\documentclass[
reprint,
groupedaddress,
showpacs,
showkeys,
amsmath, amssymb,
aps,
prx,
]{revtex4-2}

\usepackage{xcolor}
\usepackage{tabularx}
\usepackage{graphicx}% Include figure files
\usepackage{dcolumn}% Align table columns on decimal point
\usepackage{bm}% bold math
\usepackage[colorlinks=true, allcolors=blue]{hyperref}% add hypertext capabilities
\usepackage{comment}
\usepackage[normalem]{ulem}
\graphicspath{{figures/}}
\usepackage[separate-uncertainty=false]{siunitx}
\usepackage{braket}
\usepackage{xspace}
\usepackage{xstring}

\usepackage{adjustbox}
\usepackage{multirow}

\newcommand{\taub}{$\tau_{50\%}$\xspace}
\newcommand{\na}{$^{23}$Na\xspace}

\newcommand{\down}{\ensuremath{\ket{\downarrow}}\xspace}
\newcommand{\up}{\ensuremath{\ket{\uparrow}}\xspace}

\newcommand{\blue}[1]{\textcolor{blue}{}}
\newcommand{\suppress}[1]{\textcolor{brown}{}}

\newcommand*{\aref}[1]{%
	\IfBeginWith{#1}{eq:}{Eq.~\eqref{#1}}{}%r
	\IfBeginWith{#1}{fig:}{Fig.~\ref{#1}}{}%
	\IfBeginWith{#1}{tab:}{Table~\ref{#1}}{}%
	\IfBeginWith{#1}{appendix:}{Appendix~\ref{#1}}{}%
	\IfBeginWith{#1}{sec:}{Section~\ref{#1}}{}%
	}

\begin{document}

\title{Observation of false vacuum decay via bubble formation in ferromagnetic superfluids}

\date{\today}

\author{A. Zenesini$^{1,2}$}
\author{A. Berti$^1$}
\author{R. Cominotti$^1$}
\author{C. Rogora$^1$}
\author{I. G. Moss$^3$}
\author{T. P. Billam$^4$}
\author{I. Carusotto$^1$}
\author{G. Lamporesi$^{1,2}$}
\author{A. Recati$^1$}
\author{G. Ferrari$^{1,2}$}

\affiliation{$^1$Pitaevskii BEC Center, CNR-INO and Dipartimento di Fisica, Universit\`a di Trento, 38123 Trento, Italy}

\affiliation{$^2$Trento Institute for Fundamental Physics and Applications, INFN, 38123 Trento, Italy}

\affiliation{$^3$School of Mathematics, Statistics and Physics, Newcastle University, Newcastle upon Tyne, NE1 7RU, UK}

\affiliation{$^4$Joint Quantum Centre (JQC) Durham--Newcastle, School of Mathematics, Statistics and Physics, Newcastle University, Newcastle upon Tyne, NE1 7RU, UK}

\begin{abstract}
In quantum field theory, the decay of an extended metastable state into the real ground state is known as “false vacuum decay” and it takes place via the nucleation of spatially localized bubbles. Despite the large theoretical effort to estimate the nucleation rate, experimental observations were still missing. Here, we observe bubble nucleation in isolated and highly controllable superfluid atomic systems, and we find good agreement between our results, numerical simulations and instanton theory opening the way to the  emulation of out-of-equilibrium quantum field phenomena in atomic systems.
\end{abstract}

\maketitle

A supercooled gas is a classic example of a metastable state which exists just across a first order phase transition. The passage to the ground state (the liquid phase) is mediated by resonant bubble nucleation when the energy gain provided by the liquid bulk is compensated by the cost of the surface tension. This energy balance leads to a critical bubble size and a stochastic formation of the bubble typically occurs around nucleation spots given by impurities in the gas or imperfections at the container.
The extension of this idea to a quantum many-body or a quantum field system has attracted extensive attention in a wide range of scenarios and length scales, from the understanding of early universe \cite{hogan1986, Shaposhnikov:1987tw, Feeney2011} to the characterization of spin chains \cite{Lagnese21, Milsted2022}. In all these models, the metastable state at the origin of the bubble nucleation, is identified as ``false vacuum" and the role of surface tension is taken by a genuinely quantum term. In the purest form, the false vacuum decay into the ground state would take place through quantum vacuum fluctuations \cite{Coleman:1977py, Callan:1977pt} (similarly to impurities in the classical case). However, as for example in the early universe, the tunnelling is equally likely to be boosted by thermal fluctuations, and the process would be of the type styled ``vacuum decay at finite temperature" \cite{Linde:1981zj} (see \cite{Mazumdar:2018dfl,Hindmarsh2021} for a review).

In the cosmological case, the energy scales are well above any that are accessible to experiments, and the phenomenon of false vacuum decay remains one of the most important yet untested processes considered in theoretical high energy  physics. Recently, the extreme flexibility of neutral and charged atoms tabletop experiments and the advances of classical and quantum computer algorithms have paved the way for the proposal of experimental environments \cite{Fialko2015, Braden18, Billam2019, Davoudi2020, Billam2020, Billam2021, Ng2021, Song22} and virtual simulators \cite{Preskill2019WY, Abel2021}. Up to now only numerical results have been achieved and the experimental observation of an analogue to false vacuum decay would therefore be of high significance.

\begin{figure}[b]
    \centering
    \includegraphics[width = .95\columnwidth]{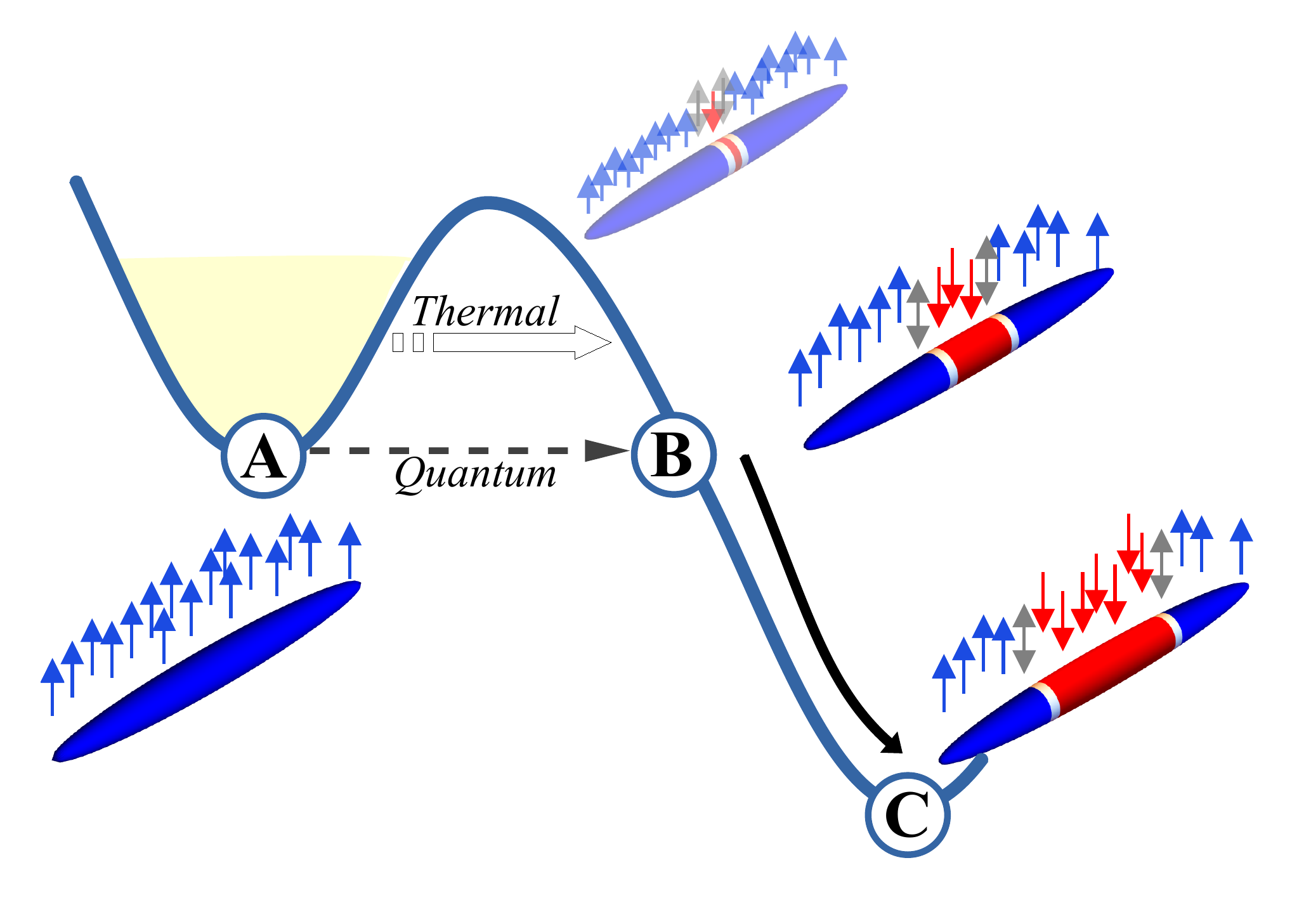}
    \caption{Mean-field energy and bubble formation. The cloud is initially prepared with all the atoms in \up  (A). While the single \down spin state is energetically lower ($E_{\downarrow}<E_{\uparrow}$) in the center of the cloud, in the low density tails the situation is opposite. The interface has a positive energy which adds up to the double minimum energy landscape emerging from the ferromagnetic interaction. Macroscopic quantum tunneling can take place resonantly to the bubble state (B) which has a \down bubble in the center, whose core energy gain compensates for the interface energy cost. The barrier crossing can be triggered by quantum fluctuations in the zero-temperature case (dashed arrow) or by thermal fluctuations at finite temperature (empty arrow). After the tunneling process, in the presence of dissipation, the bubble increases in size to reach the ground state (C), without coming back to (A).}
    \label{fig:fig1}
\end{figure}

In tabletop experiments, the observation of bubble nucleation requires several ingredients which are difficult to arrange simultaneously. First, a mean-field interaction-induced energy landscape composed of an asymmetric double well represents the minimal requirement for the decay from the metastable state to the absolute ground state via
macroscopic tunneling across the energy barrier, followed by relaxation; see sketch in \aref{fig:fig1}. Second,
unlike in the ordinary quantum tunneling of a single particle \cite{KRAMERS40,Hermann84,Grabert84}, it is an effective field describing the system that changes state. Third,
the time resolution of the experiment should cover many orders of magnitude to allow for the investigation of the predicted exponential time-dependence on the tuning parameters. This must be associated to a high stability and accuracy of the tuning parameters. An extended ferromagnetic superfluid \cite{Cominotti2022b} possesses the ideal properties to act as a field simulator, in particular its first order phase transition character, the long range coherence and the flexibility to control its experimental parameters within a stable and isolated environment. In tight analogy with supercooling, in an extended quantum system the presence of a spatial region with different magnetization to the bulk carries a positive kinetic energy due to the winding of the field at the interface, see \aref{fig:fig1}.

In this letter, we present the experimental observation of bubble formation via false vacuum decay in a quantum system. We observe that the bubble nucleation time scales exponentially with an experimental parameter that is connected to
the energy barrier properties. Theoretical and numerical simulations support our observations and allow us to confirm the quantum field origin of the decay and its thermal activation.

The experimental platform is composed of a bosonic gas of \na atoms, optically trapped and cooled below the condensation temperature. The gas is initially prepared in the internal state  $\ket{F,m_F}=\ket{1,-1}=$\down, where $F$ is the total angular momentum and $m_F$ its projection on the quantization axis. A microwave radiation with amplitude $\Omega_R$ coherently couples the \down state to $\ket{2,-2}$ =\up. The relevant scattering lengths for such a two-level system are $a_{\downarrow \downarrow}=54.5\,a_0$, $a_{\uparrow \uparrow}=64.3\,a_0, $ and $a_{\downarrow \uparrow}=54.5\,a_0$, and lead to the condition $\Delta a=(a_{\uparrow \uparrow}+a_{\downarrow \downarrow})/2 - a_{\downarrow \uparrow}<0$, i.e., to a system with a ferromagnetic ground state \cite{Cominotti2022b}.

The trapping potential is axially symmetric and harmonic in all three directions, but strongly asymmetric (axial and radial trapping frequencies $\omega_x/2\pi=\SI{20}{Hz}$ and $\omega_{\rho}/2\pi=\SI{2}{kHz}$), producing an elongated system with inhomogeneous density and spatial size given by the longitudinal and radial Thomas-Fermi radius $R_\text{x}=200\,\mu$m and $R_{\rho}=2.5\,\mu$m. 
At the end of each experimental realization, we image the two spin states independently and extract their density distributions.
The transverse confinement is tight enough to suppress the radial spin dynamics of the condensate. We therefore integrate each image along the transverse direction and obtain the integrated 1D density profiles $n_{\uparrow}(x)$ and $n_{\downarrow}(x)$, from which we extrac the profile of the relative magnetization $Z(x)=[n_{\uparrow}(x)-n_{\downarrow}(x)]/[n_{\uparrow}(x)+n_{\downarrow}(x)]$. 

\begin{figure}[t!]
    \centering
    \includegraphics[width =  \columnwidth]{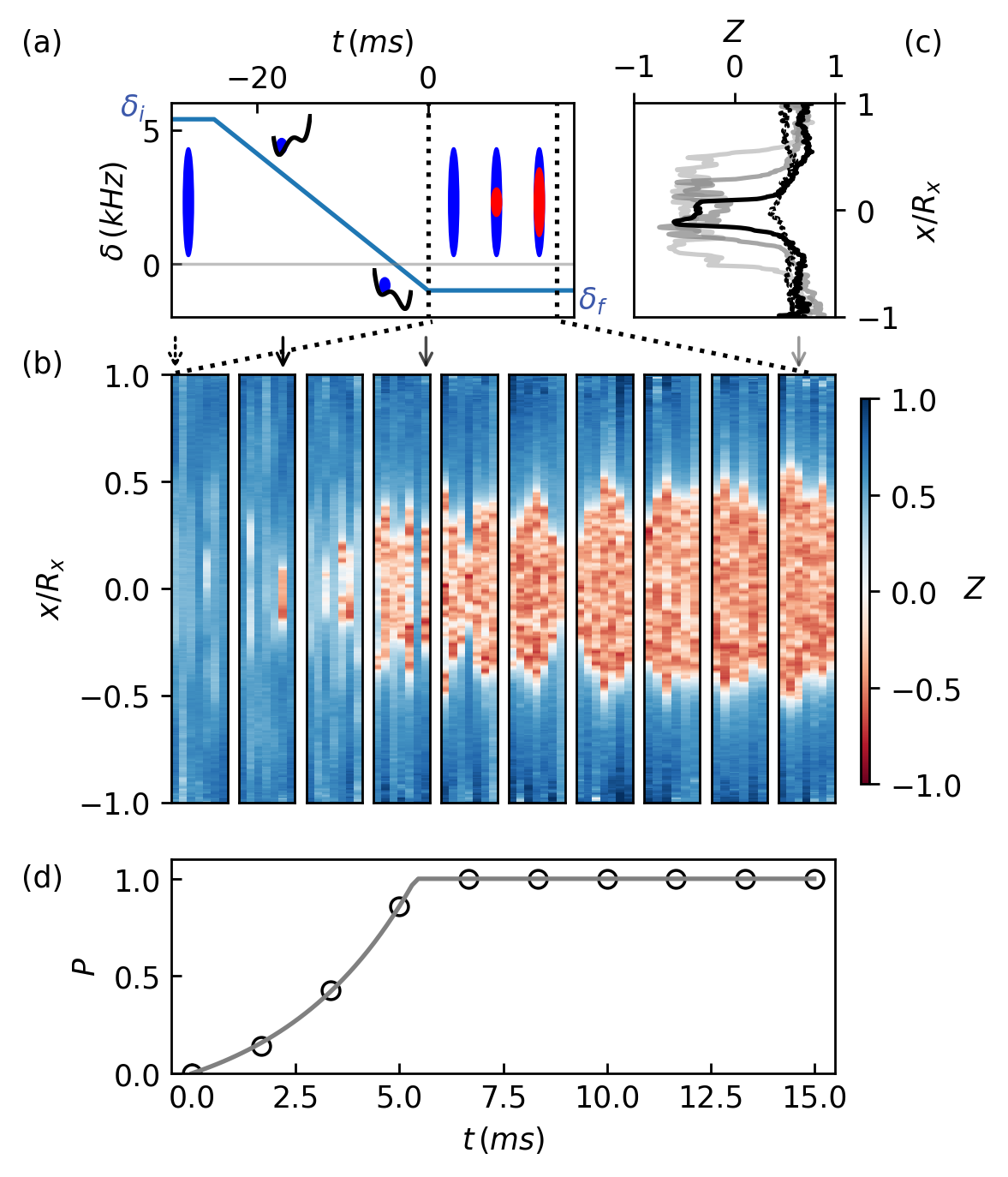}   
    \caption{ Protocols and bubble observation. a) Experimental protocol. Ellipses illustrate the cloud magnetization at different $t$ and the two sketches show the energy landscape for positive (up) and negative (down) $\delta$. b) Collection of integrated magnetization profiles $Z(x)$ after different waiting times $t$. For each value of $t$, 7 different realizations are shown. 
    c) Magnetization profiles for the realizations marked with arrows in panel (b). 
    d) Measured probability $P$ (empty circles) to observe a shot with a bubble at fixed time is shown. The probability is well fitted to an exponential curve (grey continuous line) until it saturates to 1. }
    \label{fig:fig2}
\end{figure}

The coupled two-level system can be studied by separately treating the total density ($n=n_{\uparrow}+n_{\downarrow}$) and the spin ($n_{\uparrow}-n_{\downarrow}=nZ$) degrees of freedom. While the density is simply dominated by a continuity equation, the spin degree of freedom is ruled by a magnetic mean-field Hamiltonian, which shows a first-order phase transition in the central region of the cloud for $\Omega_R<|\kappa| n$, where $\kappa\propto \Delta a$ is the relevant interaction parameter \cite{SM}.

The first-order phase transition originates from a symmetry breaking when the energy landscape as a function of the magnetization $Z$ goes from a single to a double minimum at $\Omega_R<|\kappa| n=2\pi \times 1150 \,\text{Hz}$. At fixed $\Omega_R$, the experimentally tunable parameter is the detuning $\delta$ between the two-level system and the coupling radiation.  For small enough $|\delta|$, the energy landscape $E(Z)$ is represented by an asymmetric double well, that turns symmetric for $\delta=0$. In particular, for positive $\delta$, the energy is minimized by positive values of Z, and viceversa
The relevant parameter for the bubble nucleation is the shape (height and width) of the energy barrier separating the two wells that the system needs to overcome as a field, i.e., in a macroscopic manner. This depends on $\delta$, $n$ and $\Omega_R$. When $|\delta|$ exceeds a critical value $\delta_{c}$, the metastable well disappears \cite{Cominotti2022b}. Borrowing the nomenclature from ferromagnetism, $\pm \delta_{c}$ correspond to the edges of the hysteresis region and their value depends both on $\Omega_R$ and $|\kappa| n$. 

Figure~\ref{fig:fig2}(a) illustrates the experimental protocol. We first transfer the whole system from \down to \up with a $\pi$ pulse. While keeping $\Omega_R$ constant, $\delta$ is linearly ramped down from $\delta_i/2\pi=5.5\,$kHz to a variable $\delta_f$ on a timescale between 20 and 60 ms.  
Since the ramp starts with $\delta\gg\Omega_R$, the system follows the spin rotation remaining in the local ground state until $\delta<0$ when such a local ground state becomes a metastable state; see inset in \aref{fig:fig2}(a). Once $\delta_f$ is reached, 
the states are independently imaged after a variable waiting time $t$. 

If $\delta_f>0$, the whole system is and remains in the absolute ground state \up, whereas for $\delta_f<0$, after a variable time, a macroscopic region in the central part of the system flips to \down, generating a bubble; see examples in \aref{fig:fig2}(b) and magnetization profiles in (c). On average the bubble occurrence probability is larger if the waiting time is longer [see \aref{fig:fig2}(b) and (d)].
For a quantitative analysis, at each $t$, we repeat the measurement up to 10 times in order to investigate the statistical formation of bubbles. Note that, while in uniform systems the bubbles would stochastically nucleate in random spatial positions, our nonuniform density profile of the atomic sample strongly favors the nucleation at the center of the cloud, where $\delta_f$ is closest to $\delta_c$.

\begin{figure}[t]
    \centering
    \includegraphics[width = \columnwidth]{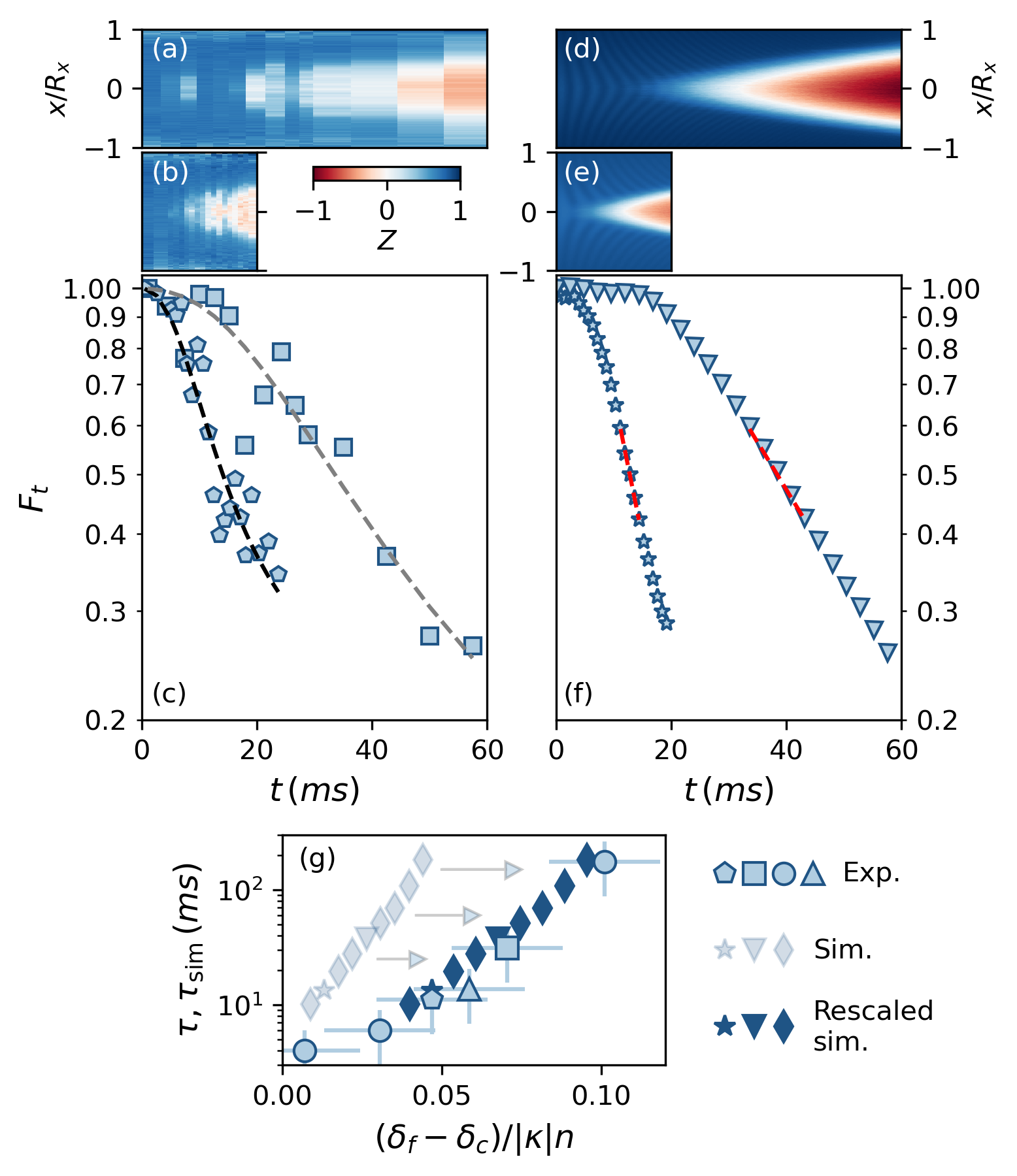}
    \caption{
    Measurement of the evolution of $Z(x)$ in time after the ramp on $\delta$ for $\Omega_R/2\pi=300\,\text{Hz}$, for $\delta_f/\Omega_R=-1.70$  in (a) and  $-1.79$ in (b). c) Value of $F_\text{t}$ evaluated in the 20 $\mu$m central region of the cloud are fitted by the empirical expression reported in the text (squares for data in (a) and pentagons for (b)). d-e) Numerical simulations for $\delta_f/\Omega_R= -1.52$ in (d) and $-1.585$ in (e). Value of $F_\text{t}$ for the simulations (triangles for data in (d) and stars for (e)). The red dashed lined are linear fits in the exponentially decaying part. g) Experimental $\tau$ and numerical $\tau_\text{sim}$ timescale of the bubble formation as a function of $(\delta_f-\delta_c)/|\kappa| n$. Error bars include statistical uncertainties on the fit and uncertainty on the $\delta_f-\delta_c$ coming from magnetic field stability and calibration. Numerical timescale of the bubble formation $\tau_\text{sim}$ is shown before (light symbols) and after (dark symbol) rescaling. 
    The empty triangle is an experimental point taken with a preparation ramp twice slower than the others, to verify the impact on the nucleation time resulting from a residual non-adiabaticity in the preparation of the sample.}
\label{fig:fig3}
\end{figure}

A useful quantity to characterize the bubble nucleation in time is $F_t=(1+\langle Z \rangle_t/\langle Z\rangle_{t=0})/2$, which was used in Ref. \cite{Lagnese21} to compare an exact diagonalization approach in a zero-temperature spin chain to instanton predictions. Here $\langle \cdot \rangle_t$ stands for  $Z$ measured at time $t$ and averaged over many realizations. 
In \aref{fig:fig3}(a) and (b), we show the average magnetization $\langle Z \rangle_t$ profile as a function of waiting time for two values of detuning. Since the bubble appears always in the center of the system, to compute $F_t$, we extract the mean magnetization $\langle Z \rangle_t$ in the central 20-$\mu$m-wide region ($\approx R_x/10$). The resulting $F_t$, plotted in panel (c), initially remains flat, and then it exponentially decays because of the bubble nucleation. Both features were also observed in Ref. \cite{Lagnese21} and the understanding of the starting plateau is still an open question from the theoretical point of view. 
We find that the measured $F_t$ is well described by the empirical function \, $(1-\epsilon)/\sqrt{1+(e^{t/\tau}-1)^2}+\epsilon$, which is 1 for $t=0$, scales as $t^2$ for small $t$ and is exponentially decaying to $\epsilon$ for large $t$. The two fitting parameters are $\tau$, that describes the characteristic timescale for the bubble formation, and $\epsilon$, that takes into account that the asymptotic magnetization $Z_{t=\infty}$ can be different from the one of the ground state, $Z_{TV}$ ($F=0$).
Note that the timescale $\tau$ is related to the exponential decay, while the empirical formula takes into account an initial plateau present in the averaged magnetisation $F_t$. (in \cite{SM} we show that the plateau length and $\tau$ are strictly connected).

Numerical simulations based on 1D Gross-Pitaevskii equations, reported in \aref{fig:fig3}(d) and (e), qualitatively reproduce the experimental observations. In the numerics, classical noise is included to simulate the effect of a finite temperature (more details can be found in \cite{SM}). Data in \aref{fig:fig3}(d) and (e) are obtained by averaging over 1000 different noisy realizations of the real-time dynamics: the large statistics allows us to directly extract the exponential decay time $\tau_\text{sim}$ through a linear fit of $\ln(F_t)$.

In \aref{fig:fig3}(g), we report six experimental values of $\tau$ obtained for $\Omega_R=2\pi \times 300$ Hz, plotted as a function of the distance from the critical detuning, $(\delta_f-\delta_c)/|\kappa| n$. The results show an exponential dependence on the tuning parameter over two orders of magnitude, from a few to hundreds of ms. Such a sensitivity to a parameter is remarkable for ultracold atoms experiments. In particular, the experimental observation of the quasi-exponential dependence of $\tau$ with respect to $\delta_f$ in an interval of the order of $\SI{100}{Hz}$ critically relies on the magnetic field stability better than a few tens of $\mu$G \cite{Farolfi19}. 

\begin{figure}[t!]
    \centering
    \includegraphics[width = .9\columnwidth]{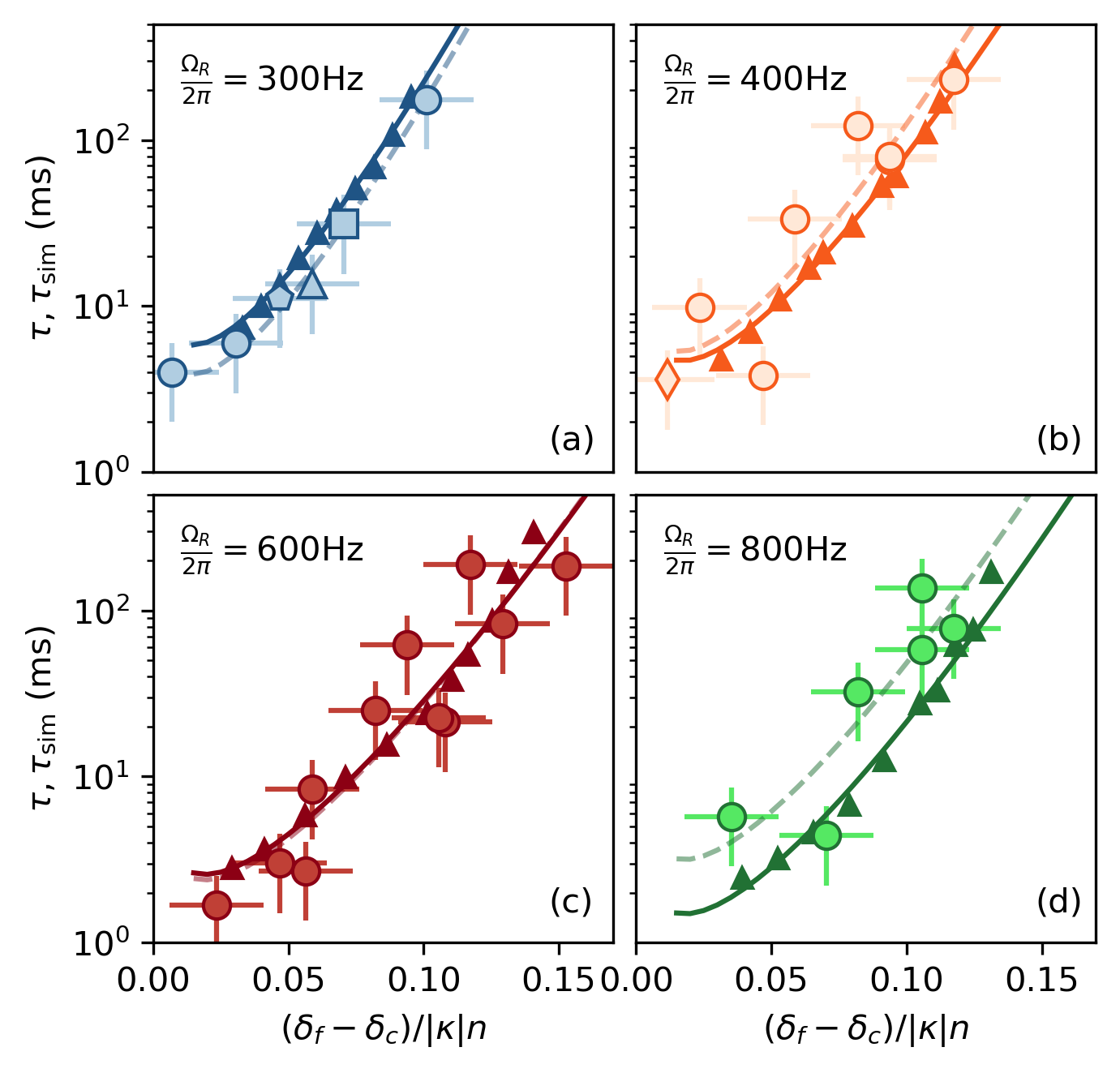}
    \caption{Decay time $\tau$ and $\tau_\text{sim}$ and instanton theory. Experimental $\tau$ and simulations $\tau_\text{sim}$ are obtained as explained in the text for $\Omega_R/2\pi= 300, 400, 600$ and $800$ Hz. A rescaling common to all $\Omega_R$ is applied to the  horizontal axes of the simulation; see text. Dashed and full curves are fits of the experimental and simulation data according to the instanton formula. Full markers stand for simulation results while empty markers for experimental data. Error bars include statistical uncertainties on the fit and uncertainty on the $\delta_f$ due to the uncertainty on the magnetic field stability.}
    \label{fig:fig4}
\end{figure}

The values of $\tau_\text{sim}$ for the simulations [light symbols in \aref{fig:fig3}(g)] qualitatively show the same behaviour of the experimental data. The agreement becomes even quantitative [dark symbols in \aref{fig:fig3}(g)], by using a rescaling of $|\kappa| n$ and a small shift of $\delta$. The need for such a rescaling was demonstrated in Ref.~\cite{Cominotti2022b}, as a consequence of dimensionality, noise and non complete adiabaticity of the preparation protocol. In \aref{fig:fig4}, we compare experimental $\tau$ and rescaled numerical $\tau_\text{sim}$, for four different values of $\Omega_R$, by using the same rescaling for all four panels. 

Our observations are consistent with the scenario of a condensate spinor field initially in a ferromagnetic metastable state, which decays via macroscopic tunneling to bubbles (domains) of the ferromagnetic ground state. The escape of a quantum field from the false vacuum, occurring via macroscopic tunneling, and the bubble formation finds a suitable description in terms of an {\it instanton},
or \textit{critical} solution to the field equations in imaginary time
\cite{Coleman:1977py, Callan:1977pt, Linde:1981zj}. Such a theory provides a threshold energy scale, below (above) which quantum (thermal) fluctuations dominate:
zero-$T$ quantum tunneling is expected to be dominant when $T$ is below the critical temperature $T^*=\hbar |\kappa| n/k_B$. Considering the peak density in our system, we estimate $T^*\simeq50$ nK. 
Although the temperature of our condensates is $T=1.5\,\mu$K $\gg T^*$, given the harmonic confinement and the exchange interaction which pushes the thermal component away from the condensate, we estimate an effective local temperature of about $250 \textrm{ nK}$ in the condensate region which is still larger than $T^*$ in the region where the bubbles appear. Therefore we expect the macroscopic tunneling to be in the thermally activated regime. 

Within the instanton approach, the bubble nucleation probability has the characteristic timescale $\tau$, which has an exponential dependence $A(E_c/k_BT)^{-1/2}e^{E_c/k_BT}$. $E_c(\delta,\kappa n,\Omega_R)$ is the energy of the critical solution and strongly depends on the shape of the many-body potential and in particular on the barrier height (\aref{fig:fig1}). The pre-factor $A$ depends on fluctuations about the critical solution, but there are very few models for which this factor is calculable, at present. We therefore regard the pre-factor $A$ as a fitting parameter in the following analysis. We can estimate $E_c$, and provide an analytical expression in the limit of vanishing metastable well (small $\delta_{f}-\delta_c$), by considering a homogeneous 1D system. The potential for the magnetization field $Z$ can be written as (see, e.g., Ref.\cite{Cominotti2022b} ) 
\begin{equation}
V(Z)= \kappa n Z^2-2\Omega(1-Z^2)^{1/2}-2\delta_{f} Z   
\end{equation}
and the instanton energy reads 
\begin{equation}
\frac{E_c}{\hbar |\kappa|n}=\sqrt{\frac{\hbar n}{2m|\kappa|}}\int_{Z_{TP}}^{Z_{FV}} \left[ \frac{ V(Z)-V(Z_{FV}) }{|\kappa|n{(1-Z^2)}} \right] ^{1/2}dZ,
\end{equation} 
where $Z_{TP}$ is the classical turning point (in the inverted potential $V$) and $Z_{F(alse)V(acuum)}$ the value of the magnetization of the metastable state. Most of our data are taken in a regime where the barrier is much smaller than the depth of the ground state well. In this limiting case the instanton energy reads \cite{SM}
\begin{equation}
    \frac{E_c}{\hbar|\kappa|n}\propto\sqrt{\frac{\hbar n}{2m|\kappa|}}\left(\frac{\delta_f-\delta_{\rm c}}{|\kappa|n}\right)^{\textstyle\frac54}
\left(\frac{\Omega_R}{|\kappa|n}\right)^{\textstyle\frac{1}{6}}
\left(\frac{|\delta_c|}{|\kappa| n}\right)^{\textstyle-\frac14},
\label{eq:anres}
\end{equation}
where $\delta_{\rm c}=\kappa n[1-(\Omega/(|\kappa|n))^{\textstyle\frac23}]^{\textstyle\frac32}$. 
We compare the previous expression to the experimental data and numerical simulations using a two-parameter fit $\ln \tau=\ln A+b \hat E_c + \ln(b\hat E_c)/2$, where $\hat E_c=\sqrt{2m|\kappa|/(\hbar n)}E_c/\hbar|\kappa|n$ is the rescaled energy. The results are shown in  Fig.~\ref{fig:fig4}. 
Considering the approximations used to derive Eq. (\ref{eq:anres}) -- in particular the absence of the trapping potential, no phase fluctuations and small barrier -- the agreement is remarkable and the instanton theory appears to capture the main dependence of the false vacuum decay rate on the microscopic parameter $\delta_f$ which is responsible for the broken $\mathbb{Z}_2$ symmetry.   

In this paper, we present solid evidence of the thermally-induced macroscopic tunneling of a coherent quantum field, manifested by bubbles of true vacuum phase nucleating in a false vacuum state. The true and false vacua are the local and global energy minimum of a ferromagnetic atomic Bose-Einstein condensate, respectively. The experimental results clearly show an exponential dependence of the decay rate on the microscopic parameters and the hysteric region width. Such a dependence is successfully captured by numerical simulations and more remarkably by a simple instanton theory based on a reduced energy functional for the magnetisation. 
Our platform paves the way to explore the process of bubble formation and growth in intricate detail, and to build a new bridge between low energy and high energy phenomena characterized by metastability within a first order phase transition. In this spirit our work opens up new avenues in the understanding of early universe, as well as ferromagnetic quantum phase transitions. 
The possibility of engineering the barrier properties via injection of tailored noise and of deterministically seeding bubbles are promising future directions for experimental investigations with focus on the role of dissipation, the existence of shortcut-to-adiabaticity \cite{Odelin2019,TORRONTEGUI2013}, the creation of entanglement, of domain wall confinement \cite{Tan21}, and relativistic and non relativistic aspects of the bubble nucleation and dynamics. 
Furthermore an experimental effort towards colder systems would allow us to reach the tunneling regime dominated by quantum fluctuations. A natural extension of the present work goes to dimensionality larger than one, where the theoretical treatment is challenging.

We thank A. Biella and P. Hauke for fruitful discussions. 
We acknowledge funding from Provincia Autonoma di Trento, from INFN through the FISH project, from the Italian MIUR under the PRIN2017 project CEnTraL (Protocol Number 20172H2SC4), from the European Union’s Horizon 2020 research and innovation Programme through the STAQS project of QuantERA II (Grant Agreement No. 101017733), from the European Research Council (ERC) under the European Union’s Horizon 2020 research and innovation programme (grant agreement No 804305), from the UK Quantum Technologies for Fundamental Physics programme (grants ST/T00584X/1 and ST/W006162/1) and from PNRR MUR project PE0000023-NQSTI. This work was supported by Q@TN, the joint lab between University of Trento, FBK - Fondazione Bruno Kessler, INFN - National Institute for Nuclear Physics and CNR - National Research Council.

%\bibliography{bibliography.bib}

\begin{thebibliography}{30}%
\makeatletter
\providecommand \@ifxundefined [1]{%
 \@ifx{#1\undefined}
}%
\providecommand \@ifnum [1]{%
 \ifnum #1\expandafter \@firstoftwo
 \else \expandafter \@secondoftwo
 \fi
}%
\providecommand \@ifx [1]{%
 \ifx #1\expandafter \@firstoftwo
 \else \expandafter \@secondoftwo
 \fi
}%
\providecommand \natexlab [1]{#1}%
\providecommand \enquote  [1]{``#1''}%
\providecommand \bibnamefont  [1]{#1}%
\providecommand \bibfnamefont [1]{#1}%
\providecommand \citenamefont [1]{#1}%
\providecommand \href@noop [0]{\@secondoftwo}%
\providecommand \href [0]{\begingroup \@sanitize@url \@href}%
\providecommand \@href[1]{\@@startlink{#1}\@@href}%
\providecommand \@@href[1]{\endgroup#1\@@endlink}%
\providecommand \@sanitize@url [0]{\catcode `\\12\catcode `\$12\catcode
  `\&12\catcode `\#12\catcode `\^12\catcode `\_12\catcode `\%12\relax}%
\providecommand \@@startlink[1]{}%
\providecommand \@@endlink[0]{}%
\providecommand \url  [0]{\begingroup\@sanitize@url \@url }%
\providecommand \@url [1]{\endgroup\@href {#1}{\urlprefix }}%
\providecommand \urlprefix  [0]{URL }%
\providecommand \Eprint [0]{\href }%
\providecommand \doibase [0]{https://doi.org/}%
\providecommand \selectlanguage [0]{\@gobble}%
\providecommand \bibinfo  [0]{\@secondoftwo}%
\providecommand \bibfield  [0]{\@secondoftwo}%
\providecommand \translation [1]{[#1]}%
\providecommand \BibitemOpen [0]{}%
\providecommand \bibitemStop [0]{}%
\providecommand \bibitemNoStop [0]{.\EOS\space}%
\providecommand \EOS [0]{\spacefactor3000\relax}%
\providecommand \BibitemShut  [1]{\csname bibitem#1\endcsname}%
\let\auto@bib@innerbib\@empty
%</preamble>
\bibitem [{\citenamefont {Hogan}(1986)}]{hogan1986}%
  \BibitemOpen
  \bibfield  {author} {\bibinfo {author} {\bibfnamefont {C.}~\bibnamefont
  {Hogan}},\ }\bibfield  {title} {\bibinfo {title} {Gravitational radiation
  from cosmological phase transitions},\ }\href@noop {} {\bibfield  {journal}
  {\bibinfo  {journal} {Monthly Notices of the Royal Astronomical Society}\
  }\textbf {\bibinfo {volume} {218}},\ \bibinfo {pages} {629} (\bibinfo {year}
  {1986})}\BibitemShut {NoStop}%
\bibitem [{\citenamefont {Shaposhnikov}(1987)}]{Shaposhnikov:1987tw}%
  \BibitemOpen
  \bibfield  {author} {\bibinfo {author} {\bibfnamefont {M.~E.}\ \bibnamefont
  {Shaposhnikov}},\ }\bibfield  {title} {\bibinfo {title} {{Baryon Asymmetry of
  the Universe in Standard Electroweak Theory}},\ }\href
  {https://doi.org/10.1016/0550-3213(87)90127-1} {\bibfield  {journal}
  {\bibinfo  {journal} {Nucl. Phys. B}\ }\textbf {\bibinfo {volume} {287}},\
  \bibinfo {pages} {757} (\bibinfo {year} {1987})}\BibitemShut {NoStop}%
\bibitem [{\citenamefont {Feeney}\ \emph {et~al.}(2011)\citenamefont {Feeney},
  \citenamefont {Johnson}, \citenamefont {Mortlock},\ and\ \citenamefont
  {Peiris}}]{Feeney2011}%
  \BibitemOpen
  \bibfield  {author} {\bibinfo {author} {\bibfnamefont {S.~M.}\ \bibnamefont
  {Feeney}}, \bibinfo {author} {\bibfnamefont {M.~C.}\ \bibnamefont {Johnson}},
  \bibinfo {author} {\bibfnamefont {D.~J.}\ \bibnamefont {Mortlock}},\ and\
  \bibinfo {author} {\bibfnamefont {H.~V.}\ \bibnamefont {Peiris}},\ }\bibfield
   {title} {\bibinfo {title} {First observational tests of eternal inflation:
  Analysis methods and wmap 7-year results},\ }\href
  {https://doi.org/10.1103/PhysRevD.84.043507} {\bibfield  {journal} {\bibinfo
  {journal} {Phys. Rev. D}\ }\textbf {\bibinfo {volume} {84}},\ \bibinfo
  {pages} {043507} (\bibinfo {year} {2011})},\ \Eprint
  {https://arxiv.org/abs/1012.3667} {arXiv:1012.3667 [astro-ph.CO]}
  \BibitemShut {NoStop}%
\bibitem [{\citenamefont {Lagnese}\ \emph {et~al.}(2021)\citenamefont
  {Lagnese}, \citenamefont {Surace}, \citenamefont {Kormos},\ and\
  \citenamefont {Calabrese}}]{Lagnese21}%
  \BibitemOpen
  \bibfield  {author} {\bibinfo {author} {\bibfnamefont {G.}~\bibnamefont
  {Lagnese}}, \bibinfo {author} {\bibfnamefont {F.~M.}\ \bibnamefont {Surace}},
  \bibinfo {author} {\bibfnamefont {M.}~\bibnamefont {Kormos}},\ and\ \bibinfo
  {author} {\bibfnamefont {P.}~\bibnamefont {Calabrese}},\ }\bibfield  {title}
  {\bibinfo {title} {False vacuum decay in quantum spin chains},\ }\href
  {https://doi.org/10.1103/PhysRevB.104.L201106} {\bibfield  {journal}
  {\bibinfo  {journal} {Phys. Rev. B}\ }\textbf {\bibinfo {volume} {104}},\
  \bibinfo {pages} {L201106} (\bibinfo {year} {2021})}\BibitemShut {NoStop}%
\bibitem [{\citenamefont {Milsted}\ \emph {et~al.}(2022)\citenamefont
  {Milsted}, \citenamefont {Liu}, \citenamefont {Preskill},\ and\ \citenamefont
  {Vidal}}]{Milsted2022}%
  \BibitemOpen
  \bibfield  {author} {\bibinfo {author} {\bibfnamefont {A.}~\bibnamefont
  {Milsted}}, \bibinfo {author} {\bibfnamefont {J.}~\bibnamefont {Liu}},
  \bibinfo {author} {\bibfnamefont {J.}~\bibnamefont {Preskill}},\ and\
  \bibinfo {author} {\bibfnamefont {G.}~\bibnamefont {Vidal}},\ }\bibfield
  {title} {\bibinfo {title} {Collisions of false-vacuum bubble walls in a
  quantum spin chain},\ }\href {https://doi.org/10.1103/PRXQuantum.3.020316}
  {\bibfield  {journal} {\bibinfo  {journal} {PRX Quantum}\ }\textbf {\bibinfo
  {volume} {3}},\ \bibinfo {pages} {020316} (\bibinfo {year}
  {2022})}\BibitemShut {NoStop}%
\bibitem [{\citenamefont {Coleman}(1977{\natexlab{a}})}]{Coleman:1977py}%
  \BibitemOpen
  \bibfield  {author} {\bibinfo {author} {\bibfnamefont {S.~R.}\ \bibnamefont
  {Coleman}},\ }\bibfield  {title} {\bibinfo {title} {{The Fate of the False
  Vacuum. 1. Semiclassical Theory}},\ }\href
  {https://doi.org/10.1103/PhysRevD.15.2929, 10.1103/PhysRevD.16.1248}
  {\bibfield  {journal} {\bibinfo  {journal} {Phys. Rev. D}\ }\textbf {\bibinfo
  {volume} {15}},\ \bibinfo {pages} {2929} (\bibinfo {year}
  {1977}{\natexlab{a}})},\ \bibinfo {note} {[Erratum: Phys. Rev. D 16, 1248
  (1977)]}\BibitemShut {NoStop}%
%%CITATION = PHRVA,D15,2929;%%
\bibitem [{\citenamefont {Callan}\ and\ \citenamefont
  {Coleman}(1977)}]{Callan:1977pt}%
  \BibitemOpen
  \bibfield  {author} {\bibinfo {author} {\bibfnamefont {C.~G.}\ \bibnamefont
  {Callan}}\ and\ \bibinfo {author} {\bibfnamefont {S.~R.}\ \bibnamefont
  {Coleman}},\ }\bibfield  {title} {\bibinfo {title} {{The Fate of the False
  Vacuum. 2. First Quantum Corrections}},\ }\href
  {https://doi.org/10.1103/PhysRevD.16.1762} {\bibfield  {journal} {\bibinfo
  {journal} {Phys. Rev. D}\ }\textbf {\bibinfo {volume} {16}},\ \bibinfo
  {pages} {1762} (\bibinfo {year} {1977})}\BibitemShut {NoStop}%
%%CITATION = PHRVA,D16,1762;%%
\bibitem [{\citenamefont {Linde}(1983)}]{Linde:1981zj}%
  \BibitemOpen
  \bibfield  {author} {\bibinfo {author} {\bibfnamefont {A.~D.}\ \bibnamefont
  {Linde}},\ }\bibfield  {title} {\bibinfo {title} {{Decay of the False Vacuum
  at Finite Temperature}},\ }\href
  {https://doi.org/10.1016/0550-3213(83)90072-X} {\bibfield  {journal}
  {\bibinfo  {journal} {Nucl. Phys. B}\ }\textbf {\bibinfo {volume} {216}},\
  \bibinfo {pages} {421} (\bibinfo {year} {1983})},\ \bibinfo {note} {[Erratum:
  Nucl.Phys.B 223, 544 (1983)]}\BibitemShut {NoStop}%
\bibitem [{\citenamefont {Mazumdar}\ and\ \citenamefont
  {White}(2019)}]{Mazumdar:2018dfl}%
  \BibitemOpen
  \bibfield  {author} {\bibinfo {author} {\bibfnamefont {A.}~\bibnamefont
  {Mazumdar}}\ and\ \bibinfo {author} {\bibfnamefont {G.}~\bibnamefont
  {White}},\ }\bibfield  {title} {\bibinfo {title} {{Review of cosmic phase
  transitions: their significance and experimental signatures}},\ }\href
  {https://doi.org/10.1088/1361-6633/ab1f55} {\bibfield  {journal} {\bibinfo
  {journal} {Rept. Prog. Phys.}\ }\textbf {\bibinfo {volume} {82}},\ \bibinfo
  {pages} {076901} (\bibinfo {year} {2019})},\ \Eprint
  {https://arxiv.org/abs/1811.01948} {arXiv:1811.01948 [hep-ph]} \BibitemShut
  {NoStop}%
\bibitem [{\citenamefont {Hindmarsh}\ \emph {et~al.}(2021)\citenamefont
  {Hindmarsh}, \citenamefont {Lüben}, \citenamefont {Lumma},\ and\
  \citenamefont {Pauly}}]{Hindmarsh2021}%
  \BibitemOpen
  \bibfield  {author} {\bibinfo {author} {\bibfnamefont {M.}~\bibnamefont
  {Hindmarsh}}, \bibinfo {author} {\bibfnamefont {M.}~\bibnamefont {Lüben}},
  \bibinfo {author} {\bibfnamefont {J.}~\bibnamefont {Lumma}},\ and\ \bibinfo
  {author} {\bibfnamefont {M.}~\bibnamefont {Pauly}},\ }\bibfield  {title}
  {\bibinfo {title} {{Phase transitions in the early universe}},\ }\href
  {https://doi.org/10.21468/SciPostPhysLectNotes.24} {\bibfield  {journal}
  {\bibinfo  {journal} {SciPost Phys. Lect. Notes}\ ,\ \bibinfo {pages} {24}}
  (\bibinfo {year} {2021})}\BibitemShut {NoStop}%
\bibitem [{\citenamefont {Fialko}\ \emph {et~al.}(2015)\citenamefont {Fialko},
  \citenamefont {Opanchuk}, \citenamefont {Sidorov}, \citenamefont {Drummond},\
  and\ \citenamefont {Brand}}]{Fialko2015}%
  \BibitemOpen
  \bibfield  {author} {\bibinfo {author} {\bibfnamefont {O.}~\bibnamefont
  {Fialko}}, \bibinfo {author} {\bibfnamefont {B.}~\bibnamefont {Opanchuk}},
  \bibinfo {author} {\bibfnamefont {A.~I.}\ \bibnamefont {Sidorov}}, \bibinfo
  {author} {\bibfnamefont {P.~D.}\ \bibnamefont {Drummond}},\ and\ \bibinfo
  {author} {\bibfnamefont {J.}~\bibnamefont {Brand}},\ }\bibfield  {title}
  {\bibinfo {title} {Fate of the false vacuum: Towards realization with
  ultra-cold atoms},\ }\href {https://doi.org/10.1209/0295-5075/110/56001}
  {\bibfield  {journal} {\bibinfo  {journal} {Europhysics Letters}\ }\textbf
  {\bibinfo {volume} {110}},\ \bibinfo {pages} {56001} (\bibinfo {year}
  {2015})}\BibitemShut {NoStop}%
\bibitem [{\citenamefont {Braden}\ \emph {et~al.}(2019)\citenamefont {Braden},
  \citenamefont {Johnson}, \citenamefont {Peiris},\ and\ \citenamefont
  {Weinfurtner}}]{Braden18}%
  \BibitemOpen
  \bibfield  {author} {\bibinfo {author} {\bibfnamefont {J.}~\bibnamefont
  {Braden}}, \bibinfo {author} {\bibfnamefont {M.~C.}\ \bibnamefont {Johnson}},
  \bibinfo {author} {\bibfnamefont {H.~V.}\ \bibnamefont {Peiris}},\ and\
  \bibinfo {author} {\bibfnamefont {S.}~\bibnamefont {Weinfurtner}},\
  }\bibfield  {title} {\bibinfo {title} {Towards the cold atom analog false
  vacuum},\ }\href {https://doi.org/doi.org/10.1007/JHEP07(2018)014} {\bibfield
   {journal} {\bibinfo  {journal} {Journal of High Energy Physics}\ }\textbf
  {\bibinfo {volume} {2018}},\ \bibinfo {pages} {2018} (\bibinfo {year}
  {2019})}\BibitemShut {NoStop}%
\bibitem [{\citenamefont {Billam}\ \emph {et~al.}(2019)\citenamefont {Billam},
  \citenamefont {Gregory}, \citenamefont {Michel},\ and\ \citenamefont
  {Moss}}]{Billam2019}%
  \BibitemOpen
  \bibfield  {author} {\bibinfo {author} {\bibfnamefont {T.~P.}\ \bibnamefont
  {Billam}}, \bibinfo {author} {\bibfnamefont {R.}~\bibnamefont {Gregory}},
  \bibinfo {author} {\bibfnamefont {F.}~\bibnamefont {Michel}},\ and\ \bibinfo
  {author} {\bibfnamefont {I.~G.}\ \bibnamefont {Moss}},\ }\bibfield  {title}
  {\bibinfo {title} {Simulating seeded vacuum decay in a cold atom system},\
  }\href {https://doi.org/10.1103/PhysRevD.100.065016} {\bibfield  {journal}
  {\bibinfo  {journal} {Phys. Rev. D}\ }\textbf {\bibinfo {volume} {100}},\
  \bibinfo {pages} {065016} (\bibinfo {year} {2019})}\BibitemShut {NoStop}%
\bibitem [{\citenamefont {Davoudi}\ \emph {et~al.}(2020)\citenamefont
  {Davoudi}, \citenamefont {Hafezi}, \citenamefont {Monroe}, \citenamefont
  {Pagano}, \citenamefont {Seif},\ and\ \citenamefont {Shaw}}]{Davoudi2020}%
  \BibitemOpen
  \bibfield  {author} {\bibinfo {author} {\bibfnamefont {Z.}~\bibnamefont
  {Davoudi}}, \bibinfo {author} {\bibfnamefont {M.}~\bibnamefont {Hafezi}},
  \bibinfo {author} {\bibfnamefont {C.}~\bibnamefont {Monroe}}, \bibinfo
  {author} {\bibfnamefont {G.}~\bibnamefont {Pagano}}, \bibinfo {author}
  {\bibfnamefont {A.}~\bibnamefont {Seif}},\ and\ \bibinfo {author}
  {\bibfnamefont {A.}~\bibnamefont {Shaw}},\ }\bibfield  {title} {\bibinfo
  {title} {Towards analog quantum simulations of lattice gauge theories with
  trapped ions},\ }\href {https://doi.org/10.1103/PhysRevResearch.2.023015}
  {\bibfield  {journal} {\bibinfo  {journal} {Phys. Rev. Res.}\ }\textbf
  {\bibinfo {volume} {2}},\ \bibinfo {pages} {023015} (\bibinfo {year}
  {2020})}\BibitemShut {NoStop}%
\bibitem [{\citenamefont {Billam}\ \emph {et~al.}(2020)\citenamefont {Billam},
  \citenamefont {Brown},\ and\ \citenamefont {Moss}}]{Billam2020}%
  \BibitemOpen
  \bibfield  {author} {\bibinfo {author} {\bibfnamefont {T.~P.}\ \bibnamefont
  {Billam}}, \bibinfo {author} {\bibfnamefont {K.}~\bibnamefont {Brown}},\ and\
  \bibinfo {author} {\bibfnamefont {I.~G.}\ \bibnamefont {Moss}},\ }\bibfield
  {title} {\bibinfo {title} {Simulating cosmological supercooling with a
  cold-atom system},\ }\href {https://doi.org/10.1103/PhysRevA.102.043324}
  {\bibfield  {journal} {\bibinfo  {journal} {Phys. Rev. A}\ }\textbf {\bibinfo
  {volume} {102}},\ \bibinfo {pages} {043324} (\bibinfo {year}
  {2020})}\BibitemShut {NoStop}%
\bibitem [{\citenamefont {Billam}\ \emph {et~al.}(2021)\citenamefont {Billam},
  \citenamefont {Brown}, \citenamefont {Groszek},\ and\ \citenamefont
  {Moss}}]{Billam2021}%
  \BibitemOpen
  \bibfield  {author} {\bibinfo {author} {\bibfnamefont {T.~P.}\ \bibnamefont
  {Billam}}, \bibinfo {author} {\bibfnamefont {K.}~\bibnamefont {Brown}},
  \bibinfo {author} {\bibfnamefont {A.~J.}\ \bibnamefont {Groszek}},\ and\
  \bibinfo {author} {\bibfnamefont {I.~G.}\ \bibnamefont {Moss}},\ }\bibfield
  {title} {\bibinfo {title} {Simulating cosmological supercooling with a cold
  atom system. ii. thermal damping and parametric instability},\ }\href
  {https://doi.org/10.1103/PhysRevA.104.053309} {\bibfield  {journal} {\bibinfo
   {journal} {Phys. Rev. A}\ }\textbf {\bibinfo {volume} {104}},\ \bibinfo
  {pages} {053309} (\bibinfo {year} {2021})}\BibitemShut {NoStop}%
\bibitem [{\citenamefont {Ng}\ \emph {et~al.}(2021)\citenamefont {Ng},
  \citenamefont {Opanchuk}, \citenamefont {Thenabadu}, \citenamefont {Reid},\
  and\ \citenamefont {Drummond}}]{Ng2021}%
  \BibitemOpen
  \bibfield  {author} {\bibinfo {author} {\bibfnamefont {K.~L.}\ \bibnamefont
  {Ng}}, \bibinfo {author} {\bibfnamefont {B.}~\bibnamefont {Opanchuk}},
  \bibinfo {author} {\bibfnamefont {M.}~\bibnamefont {Thenabadu}}, \bibinfo
  {author} {\bibfnamefont {M.}~\bibnamefont {Reid}},\ and\ \bibinfo {author}
  {\bibfnamefont {P.~D.}\ \bibnamefont {Drummond}},\ }\bibfield  {title}
  {\bibinfo {title} {Fate of the false vacuum: Finite temperature, entropy, and
  topological phase in quantum simulations of the early universe},\ }\href
  {https://doi.org/10.1103/PRXQuantum.2.010350} {\bibfield  {journal} {\bibinfo
   {journal} {PRX Quantum}\ }\textbf {\bibinfo {volume} {2}},\ \bibinfo {pages}
  {010350} (\bibinfo {year} {2021})}\BibitemShut {NoStop}%
\bibitem [{\citenamefont {Song}\ \emph {et~al.}(2022)\citenamefont {Song},
  \citenamefont {Dutta}, \citenamefont {Bhave}, \citenamefont {Yu},
  \citenamefont {Carter}, \citenamefont {Cooper},\ and\ \citenamefont
  {Schneider}}]{Song22}%
  \BibitemOpen
  \bibfield  {author} {\bibinfo {author} {\bibfnamefont {B.}~\bibnamefont
  {Song}}, \bibinfo {author} {\bibfnamefont {S.}~\bibnamefont {Dutta}},
  \bibinfo {author} {\bibfnamefont {S.}~\bibnamefont {Bhave}}, \bibinfo
  {author} {\bibfnamefont {J.-C.}\ \bibnamefont {Yu}}, \bibinfo {author}
  {\bibfnamefont {E.}~\bibnamefont {Carter}}, \bibinfo {author} {\bibfnamefont
  {N.}~\bibnamefont {Cooper}},\ and\ \bibinfo {author} {\bibfnamefont
  {U.}~\bibnamefont {Schneider}},\ }\bibfield  {title} {\bibinfo {title}
  {{Realizing discontinuous quantum phase transitions in a strongly correlated
  driven optical lattice}},\ }\href
  {https://doi.org/10.1038/s41567-021-01476-w} {\bibfield  {journal} {\bibinfo
  {journal} {Nature Physics}\ }\textbf {\bibinfo {volume} {18}},\ \bibinfo
  {pages} {259–264} (\bibinfo {year} {2022})}\BibitemShut {NoStop}%
\bibitem [{\citenamefont {Preskill}(2019)}]{Preskill2019WY}%
  \BibitemOpen
  \bibfield  {author} {\bibinfo {author} {\bibfnamefont {J.}~\bibnamefont
  {Preskill}},\ }\bibfield  {title} {\bibinfo {title} {{Simulating quantum
  field theory with a quantum computer}},\ }\href
  {https://doi.org/10.22323/1.334.0024} {\bibfield  {journal} {\bibinfo
  {journal} {PoS}\ }\textbf {\bibinfo {volume} {LATTICE2018}},\ \bibinfo
  {pages} {024} (\bibinfo {year} {2019})}\BibitemShut {NoStop}%
\bibitem [{\citenamefont {Abel}\ and\ \citenamefont
  {Spannowsky}(2021)}]{Abel2021}%
  \BibitemOpen
  \bibfield  {author} {\bibinfo {author} {\bibfnamefont {S.}~\bibnamefont
  {Abel}}\ and\ \bibinfo {author} {\bibfnamefont {M.}~\bibnamefont
  {Spannowsky}},\ }\bibfield  {title} {\bibinfo {title}
  {Quantum-field-theoretic simulation platform for observing the fate of the
  false vacuum},\ }\href {https://doi.org/10.1103/PRXQuantum.2.010349}
  {\bibfield  {journal} {\bibinfo  {journal} {PRX Quantum}\ }\textbf {\bibinfo
  {volume} {2}},\ \bibinfo {pages} {010349} (\bibinfo {year}
  {2021})}\BibitemShut {NoStop}%
\bibitem [{\citenamefont {Kramers}(1940)}]{KRAMERS40}%
  \BibitemOpen
  \bibfield  {author} {\bibinfo {author} {\bibfnamefont {H.}~\bibnamefont
  {Kramers}},\ }\bibfield  {title} {\bibinfo {title} {Brownian motion in a
  field of force and the diffusion model of chemical reactions},\ }\href
  {https://doi.org/https://doi.org/10.1016/S0031-8914(40)90098-2} {\bibfield
  {journal} {\bibinfo  {journal} {Physica}\ }\textbf {\bibinfo {volume} {7}},\
  \bibinfo {pages} {284} (\bibinfo {year} {1940})}\BibitemShut {NoStop}%
\bibitem [{\citenamefont {Grabert}\ and\ \citenamefont
  {Weiss}(1984)}]{Hermann84}%
  \BibitemOpen
  \bibfield  {author} {\bibinfo {author} {\bibfnamefont {H.}~\bibnamefont
  {Grabert}}\ and\ \bibinfo {author} {\bibfnamefont {U.}~\bibnamefont
  {Weiss}},\ }\bibfield  {title} {\bibinfo {title} {Crossover from thermal
  hopping to quantum tunneling},\ }\href
  {https://doi.org/10.1103/PhysRevLett.53.1787} {\bibfield  {journal} {\bibinfo
   {journal} {Phys. Rev. Lett.}\ }\textbf {\bibinfo {volume} {53}},\ \bibinfo
  {pages} {1787} (\bibinfo {year} {1984})}\BibitemShut {NoStop}%
\bibitem [{\citenamefont {Grabert}\ \emph {et~al.}(1984)\citenamefont
  {Grabert}, \citenamefont {Weiss},\ and\ \citenamefont {Hanggi}}]{Grabert84}%
  \BibitemOpen
  \bibfield  {author} {\bibinfo {author} {\bibfnamefont {H.}~\bibnamefont
  {Grabert}}, \bibinfo {author} {\bibfnamefont {U.}~\bibnamefont {Weiss}},\
  and\ \bibinfo {author} {\bibfnamefont {P.}~\bibnamefont {Hanggi}},\
  }\bibfield  {title} {\bibinfo {title} {Quantum tunneling in dissipative
  systems at finite temperatures},\ }\href
  {https://doi.org/10.1103/PhysRevLett.52.2193} {\bibfield  {journal} {\bibinfo
   {journal} {Phys. Rev. Lett.}\ }\textbf {\bibinfo {volume} {52}},\ \bibinfo
  {pages} {2193} (\bibinfo {year} {1984})}\BibitemShut {NoStop}%
\bibitem [{\citenamefont {Cominotti}\ \emph {et~al.}()\citenamefont
  {Cominotti}, \citenamefont {Berti}, \citenamefont {Dulin}, \citenamefont
  {Rogora}, \citenamefont {Lamporesi}, \citenamefont {Carusotto}, \citenamefont
  {Recati}, \citenamefont {Zenesini},\ and\ \citenamefont
  {Ferrari}}]{Cominotti2022b}%
  \BibitemOpen
  \bibfield  {author} {\bibinfo {author} {\bibfnamefont {R.}~\bibnamefont
  {Cominotti}}, \bibinfo {author} {\bibfnamefont {A.}~\bibnamefont {Berti}},
  \bibinfo {author} {\bibfnamefont {C.}~\bibnamefont {Dulin}}, \bibinfo
  {author} {\bibfnamefont {C.}~\bibnamefont {Rogora}}, \bibinfo {author}
  {\bibfnamefont {G.}~\bibnamefont {Lamporesi}}, \bibinfo {author}
  {\bibfnamefont {I.}~\bibnamefont {Carusotto}}, \bibinfo {author}
  {\bibfnamefont {A.}~\bibnamefont {Recati}}, \bibinfo {author} {\bibfnamefont
  {A.}~\bibnamefont {Zenesini}},\ and\ \bibinfo {author} {\bibfnamefont
  {G.}~\bibnamefont {Ferrari}},\ }\href@noop {} {\bibinfo {title}
  {Ferromagnetism in an extended coherently-coupled atomic superfluid. (to be
  published in $\text{Phys. Rev. X, 2023}$)}},\ \Eprint
  {https://arxiv.org/abs/2209.13235} {arXiv:2209.13235 [cond-mat.quant-gas]}
  \BibitemShut {NoStop}%
\bibitem [{SM()}]{SM}%
  \BibitemOpen
  \href@noop {} {\bibinfo {title} {{See Supplemental Material [url] for
  additional information on experimental calibrations and theoretical
  models}}}\BibitemShut {NoStop}%
\bibitem [{\citenamefont {Farolfi}\ \emph {et~al.}(2019)\citenamefont
  {Farolfi}, \citenamefont {Trypogeorgos}, \citenamefont {Colzi}, \citenamefont
  {Fava}, \citenamefont {Lamporesi},\ and\ \citenamefont
  {Ferrari}}]{Farolfi19}%
  \BibitemOpen
  \bibfield  {author} {\bibinfo {author} {\bibfnamefont {A.}~\bibnamefont
  {Farolfi}}, \bibinfo {author} {\bibfnamefont {D.}~\bibnamefont
  {Trypogeorgos}}, \bibinfo {author} {\bibfnamefont {G.}~\bibnamefont {Colzi}},
  \bibinfo {author} {\bibfnamefont {E.}~\bibnamefont {Fava}}, \bibinfo {author}
  {\bibfnamefont {G.}~\bibnamefont {Lamporesi}},\ and\ \bibinfo {author}
  {\bibfnamefont {G.}~\bibnamefont {Ferrari}},\ }\bibfield  {title} {\bibinfo
  {title} {{Design and characterization of a compact magnetic shield for
  ultracold atomic gas experiments}},\ }\href
  {https://doi.org/10.1063/1.5119915} {\bibfield  {journal} {\bibinfo
  {journal} {Review of Scientific Instruments}\ }\textbf {\bibinfo {volume}
  {90}},\ \bibinfo {pages} {115114} (\bibinfo {year} {2019})}\BibitemShut
  {NoStop}%
\bibitem [{\citenamefont {Gu\'ery-Odelin}\ \emph {et~al.}(2019)\citenamefont
  {Gu\'ery-Odelin}, \citenamefont {Ruschhaupt}, \citenamefont {Kiely},
  \citenamefont {Torrontegui}, \citenamefont {Mart\'{\i}nez-Garaot},\ and\
  \citenamefont {Muga}}]{Odelin2019}%
  \BibitemOpen
  \bibfield  {author} {\bibinfo {author} {\bibfnamefont {D.}~\bibnamefont
  {Gu\'ery-Odelin}}, \bibinfo {author} {\bibfnamefont {A.}~\bibnamefont
  {Ruschhaupt}}, \bibinfo {author} {\bibfnamefont {A.}~\bibnamefont {Kiely}},
  \bibinfo {author} {\bibfnamefont {E.}~\bibnamefont {Torrontegui}}, \bibinfo
  {author} {\bibfnamefont {S.}~\bibnamefont {Mart\'{\i}nez-Garaot}},\ and\
  \bibinfo {author} {\bibfnamefont {J.~G.}\ \bibnamefont {Muga}},\ }\bibfield
  {title} {\bibinfo {title} {Shortcuts to adiabaticity: Concepts, methods, and
  applications},\ }\href {https://doi.org/10.1103/RevModPhys.91.045001}
  {\bibfield  {journal} {\bibinfo  {journal} {Rev. Mod. Phys.}\ }\textbf
  {\bibinfo {volume} {91}},\ \bibinfo {pages} {045001} (\bibinfo {year}
  {2019})}\BibitemShut {NoStop}%
\bibitem [{\citenamefont {Torrontegui}\ \emph {et~al.}(2013)\citenamefont
  {Torrontegui}, \citenamefont {Ibáñez}, \citenamefont {Martínez-Garaot},
  \citenamefont {Modugno}, \citenamefont {{del Campo}}, \citenamefont
  {Guéry-Odelin}, \citenamefont {Ruschhaupt}, \citenamefont {Chen},\ and\
  \citenamefont {Muga}}]{TORRONTEGUI2013}%
  \BibitemOpen
  \bibfield  {author} {\bibinfo {author} {\bibfnamefont {E.}~\bibnamefont
  {Torrontegui}}, \bibinfo {author} {\bibfnamefont {S.}~\bibnamefont
  {Ibáñez}}, \bibinfo {author} {\bibfnamefont {S.}~\bibnamefont
  {Martínez-Garaot}}, \bibinfo {author} {\bibfnamefont {M.}~\bibnamefont
  {Modugno}}, \bibinfo {author} {\bibfnamefont {A.}~\bibnamefont {{del
  Campo}}}, \bibinfo {author} {\bibfnamefont {D.}~\bibnamefont
  {Guéry-Odelin}}, \bibinfo {author} {\bibfnamefont {A.}~\bibnamefont
  {Ruschhaupt}}, \bibinfo {author} {\bibfnamefont {X.}~\bibnamefont {Chen}},\
  and\ \bibinfo {author} {\bibfnamefont {J.~G.}\ \bibnamefont {Muga}},\
  }\bibfield  {title} {\bibinfo {title} {Chapter 2 - shortcuts to
  adiabaticity},\ }in\ \href
  {https://doi.org/https://doi.org/10.1016/B978-0-12-408090-4.00002-5} {\emph
  {\bibinfo {booktitle} {Advances in Atomic, Molecular, and Optical
  Physics}}},\ \bibinfo {series} {Advances In Atomic, Molecular, and Optical
  Physics}, Vol.~\bibinfo {volume} {62},\ \bibinfo {editor} {edited by\
  \bibinfo {editor} {\bibfnamefont {E.}~\bibnamefont {Arimondo}}, \bibinfo
  {editor} {\bibfnamefont {P.~R.}\ \bibnamefont {Berman}},\ and\ \bibinfo
  {editor} {\bibfnamefont {C.~C.}\ \bibnamefont {Lin}}}\ (\bibinfo  {publisher}
  {Academic Press},\ \bibinfo {year} {2013})\ pp.\ \bibinfo {pages}
  {117--169}\BibitemShut {NoStop}%
\bibitem [{\citenamefont {Tan}\ \emph {et~al.}(2021)\citenamefont {Tan},
  \citenamefont {Becker}, \citenamefont {Liu}, \citenamefont {Pagano},
  \citenamefont {Collins}, \citenamefont {De}, \citenamefont {Feng},
  \citenamefont {Kaplan}, \citenamefont {Kyprianidis}, \citenamefont
  {Lundgren}, \citenamefont {Morong}, \citenamefont {Whitsitt}, \citenamefont
  {Gorshkov},\ and\ \citenamefont {Monroe}}]{Tan21}%
  \BibitemOpen
  \bibfield  {author} {\bibinfo {author} {\bibfnamefont {W.~L.}\ \bibnamefont
  {Tan}}, \bibinfo {author} {\bibfnamefont {P.}~\bibnamefont {Becker}},
  \bibinfo {author} {\bibfnamefont {F.}~\bibnamefont {Liu}}, \bibinfo {author}
  {\bibfnamefont {G.}~\bibnamefont {Pagano}}, \bibinfo {author} {\bibfnamefont
  {K.~S.}\ \bibnamefont {Collins}}, \bibinfo {author} {\bibfnamefont
  {A.}~\bibnamefont {De}}, \bibinfo {author} {\bibfnamefont {L.}~\bibnamefont
  {Feng}}, \bibinfo {author} {\bibfnamefont {H.~B.}\ \bibnamefont {Kaplan}},
  \bibinfo {author} {\bibfnamefont {A.}~\bibnamefont {Kyprianidis}}, \bibinfo
  {author} {\bibfnamefont {R.}~\bibnamefont {Lundgren}}, \bibinfo {author}
  {\bibfnamefont {W.}~\bibnamefont {Morong}}, \bibinfo {author} {\bibfnamefont
  {S.}~\bibnamefont {Whitsitt}}, \bibinfo {author} {\bibfnamefont {A.~V.}\
  \bibnamefont {Gorshkov}},\ and\ \bibinfo {author} {\bibfnamefont
  {C.}~\bibnamefont {Monroe}},\ }\bibfield  {title} {\bibinfo {title}
  {Domain-wall confinement and dynamics in a quantum simulator},\ }\href
  {https://doi.org/0.1038/s41567-021-01194-3} {\bibfield  {journal} {\bibinfo
  {journal} {Nature Physics}\ }\textbf {\bibinfo {volume} {17}},\ \bibinfo
  {pages} {742–} (\bibinfo {year} {2021})}\BibitemShut {NoStop}%
\bibitem [{\citenamefont {Coleman}(1977{\natexlab{b}})}]{Coleman1977a}%
  \BibitemOpen
  \bibfield  {author} {\bibinfo {author} {\bibfnamefont {S.}~\bibnamefont
  {Coleman}},\ }\bibfield  {title} {\bibinfo {title} {Fate of the false vacuum:
  Semiclassical theory},\ }\href {https://doi.org/10.1103/PhysRevD.15.2929}
  {\bibfield  {journal} {\bibinfo  {journal} {Phys. Rev. D}\ }\textbf {\bibinfo
  {volume} {15}},\ \bibinfo {pages} {2929} (\bibinfo {year}
  {1977}{\natexlab{b}})}\BibitemShut {NoStop}%
\end{thebibliography}


\onecolumngrid

\newpage

\renewcommand{\thefigure}{S\arabic{figure}}
\setcounter{figure}{0}
\section*{Supplementary Material}

\section{Ferromagnetism in elongated mixtures}
\label{ferro}

The ferromagnetic properties of atomic superfluid coupled mixtures are experimentally measured and discussed in \cite{Cominotti2022b}. Here we summarize the key ingredients which help understanding the results presented in the main text of the article.

Our system is composed of two sodium hyperfine states $\ket{F,m_F}=\ket{2,-2}\equiv\up$ and $\ket{1,-1}\equiv\down$, where $F$ is the total angular momentum and $m_F$ its projection.
The two populations $n_\uparrow(x,y)$ and $n_\downarrow(x,y)$ are independently measured by shadow imaging. Starting from the two two-dimensional pictures of the cloud, we determine the relative magnetization $Z(x)$ as $Z(x) = (n_\uparrow(x)-n_\downarrow(x))/n(x)$, where $n_{\uparrow,(\downarrow)}(x)=\int n_{\uparrow,(\downarrow)}(x,y)dy$ and $n(x)=\int(n_\uparrow(x,y)+n_\downarrow(x,y))dy$ are the 1D integrated densities. The integration along $y$ takes advantage of the suppressed radial dynamics.
In local density approximation (LDA), the energy per particle associated to the spin channel of the mixture is 
\begin{equation}
     E(Z, \phi) \propto  - \frac{\delta_f}{2} Z + \frac {\kappa n}{2} Z^2 - \Omega_\text{R} \sqrt{1-Z^2}\cos\phi \label{eq:energy2}
\end{equation}
where the phase $\phi$ is the relative phase between $\up$ and $\down$. The detuning $\delta_f$ used in the text is equal to $\delta_\mathrm{B}+n \Delta$ where $\delta_\mathrm{B}$ is the experimental controllable detuning. The quantity $\kappa$ and $\Delta$ are associated to the collisional proprieties of the mixture and are
\begin{align}
    \Delta &\equiv \frac{g_{\downarrow\downarrow}-g_{\uparrow\uparrow}}{2 \hbar} <0  \\
    \kappa &\equiv \frac{g_{\downarrow\downarrow}+g_{\uparrow\uparrow}}{2\hbar}-\frac{g_{\downarrow\uparrow}}{\hbar} <0
\end{align} where $g_{\downarrow\downarrow}, g_{\uparrow\uparrow}$ and $g_{\downarrow\uparrow}$ are the two intra species and the inter species coupling constants. Note that $n \Delta$ derives from the \up and \down self interaction asymmetry.

\begin{figure}[b!]
    \centering
    \includegraphics[width = 0.75\columnwidth]{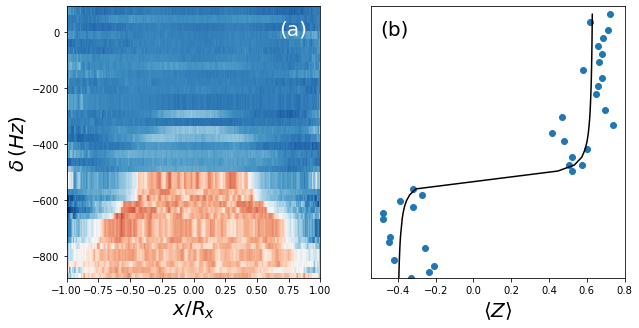}
    \caption{Determination of $\delta_c$ from the hysteresis end. a) Integrated profile of the magnetization $Z(x)$ for different  the detuning $\delta$ and no waiting time. b) The change of sign of the magnetization $\langle Z \rangle$ averaged on the central 40 pixels allows to determine $\delta_c=-517(17)\,\mathrm{Hz}$, where the uncertainties comes from the statistical uncertainties on the sigmoidal fit (orange line). This calibration is used for a subset of data shown in the main text for $\Omega_R/(2\pi)=400$\,Hz.}
    \label{S1}
\end{figure}

In an elongated cloud having a parabolic Thomas Fermi density profile, the ferromagnetic phase is located in the center of the cloud where the non liner term $ |\kappa| n Z^2 /2$ is maximal. Under the condition $|\kappa| n<\Omega$, the energy per particle is characterized by a symmetric double minimum structures a signature of the symmetry breaking typical of the ferromagnetic phase. At non zero detuning, the symmetry of the two wells is broken. Thanks to the tuning knob $\delta_B$, which is linearly proportional to the applied magnetic field, one can change the relative energy difference between the two energy minima, converting one or the other state into the absolute ground state or the metastable state. The tails of the cloud remain in the paramagnetic regime, having smaller density, and $Z$ of the only energy minimum is unambiguously determined by $\delta_\text{B}$.

Due to the asymmetry between $\up$ and $\down$, there exists a range of values of $\delta_\text{B}$ where the sign of the $Z$ at the energy minima in the center ($-$) and at the tails ($+$) is opposite, but the system can still maintain a homogeneous positively-magnetized profile being metastable in the center. When the detuning is decreased below the critical value $\delta_c$ (see main text), the metastable minimum disappears resulting in a unique steady magnetic profile with negative $Z$ in the center and positive $Z$ in the tails.

While the spin energy profiles of \aref{eq:energy2} are useful to explain the presence of two minima separated, this LDA representation only shows the LDA energy landscape per particle and not the total energy of the system. For instance, the LDA energy profiles don't include the contribution coming from the interface between opposite $Z$, whose kinetic energy represents a further contribution to the total energy barrier, as intended to be shown in Fig.\,1 in the main text.

\section{Calibration and analysis procedure}

An important calibration concerns the determination of the critical detuning at which the double well energy landscape is expected to disappear. We determine $\delta_c$ by performing the same protocols used in \cite{Cominotti2022b}  to measure the hysteresis width of the ferromagnetic regime. This consists in the same ramp shown Fig.\,2(a) of the main text, applied with a null waiting time. Figure S.\ref{S1}(a) shows the integrated magnetization $Z(x)$ as a function of the  final value of $\delta_f$. In the case presented in figure, we identify $\delta_c/2\pi=710(19)$ Hz by fitting the magnetisation averaged over the central 40 pixel with a sigmoidal function, see Fig.\,\ref{S1}.b.

The data used in the main text are obtained in the range of $\delta$ directly above the critical one. Thanks to the appearing of the bubble  in the center of the cloud, we first determine the presence of the bubble by fixing a threshold $Z_\mathrm{bubble}=0.2$. If the average magnetization in the central 40 pixels is below $Z_\mathrm{bubble}$, one bubble is counted. The total bubble counts at fixed waiting time determines the probability $P$, as plotted in Fig.\,2(c) of the main text. We verify that the choice of the threshold $Z_\mathrm{bubble}$ and the averaging area do not critically impact on the outcomes presented here.

\begin{figure}
    \centering
    \includegraphics[width = .65\columnwidth]{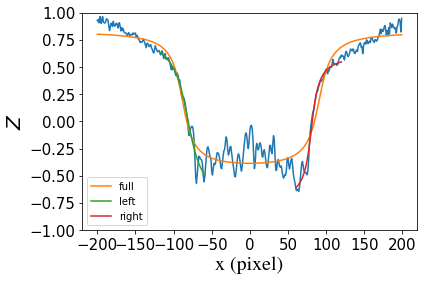}
    \caption{Magnetization profile of a typical experimental shot showing the appearance of a bubble. After a first overall fit (orange line), two independent fits are performed at the two interfaces (left green and right red).}
    \label{S2}
\end{figure}

Once the bubble is detected, the full magnetization profile is initially fitted by using a double sigmoidal function, 
\begin{equation}
    A \left [\arctan \left (\frac{x-x_r}{s_r} \right)-\arctan \left ( \frac{x-x_l}{s_l} \right )\right ]
\end{equation}
where $A$ is the amplitude and $x_{(r),[l]}$ and $s_{(r),[l]}$ are the (right) [left] centers and sigmas of the two sigmoids, see orange line in Fig.S2.
The positions $x_{(l),[r]}$ are then used as starting values for a second fitting routine that independently analyses the left and right bubble interfaces. This second step is used to better determine the exact positions of the interfaces without the effects of cloud asymmetry and offsets. The obtained values $x_{(l),[r]}$ allow to determine the bubble size as $\sigma_x=x_r-x_l$

\section{Determination of $\tau$ and alternative \taub}
\label{tau50}
In the main text we explain how we determine the characteristic decay time $\tau$ by fitting $F_t$ to $(1-\epsilon)/\sqrt{1+(e^{t/\tau}-1)^2}+\epsilon$. This formula allows us to extract $\tau$ even for experimental sequences with limited statistics and it results to be robust against the initialisation of the fitting parameters.

\begin{figure}
    \centering
    \includegraphics[width = 0.75\columnwidth]{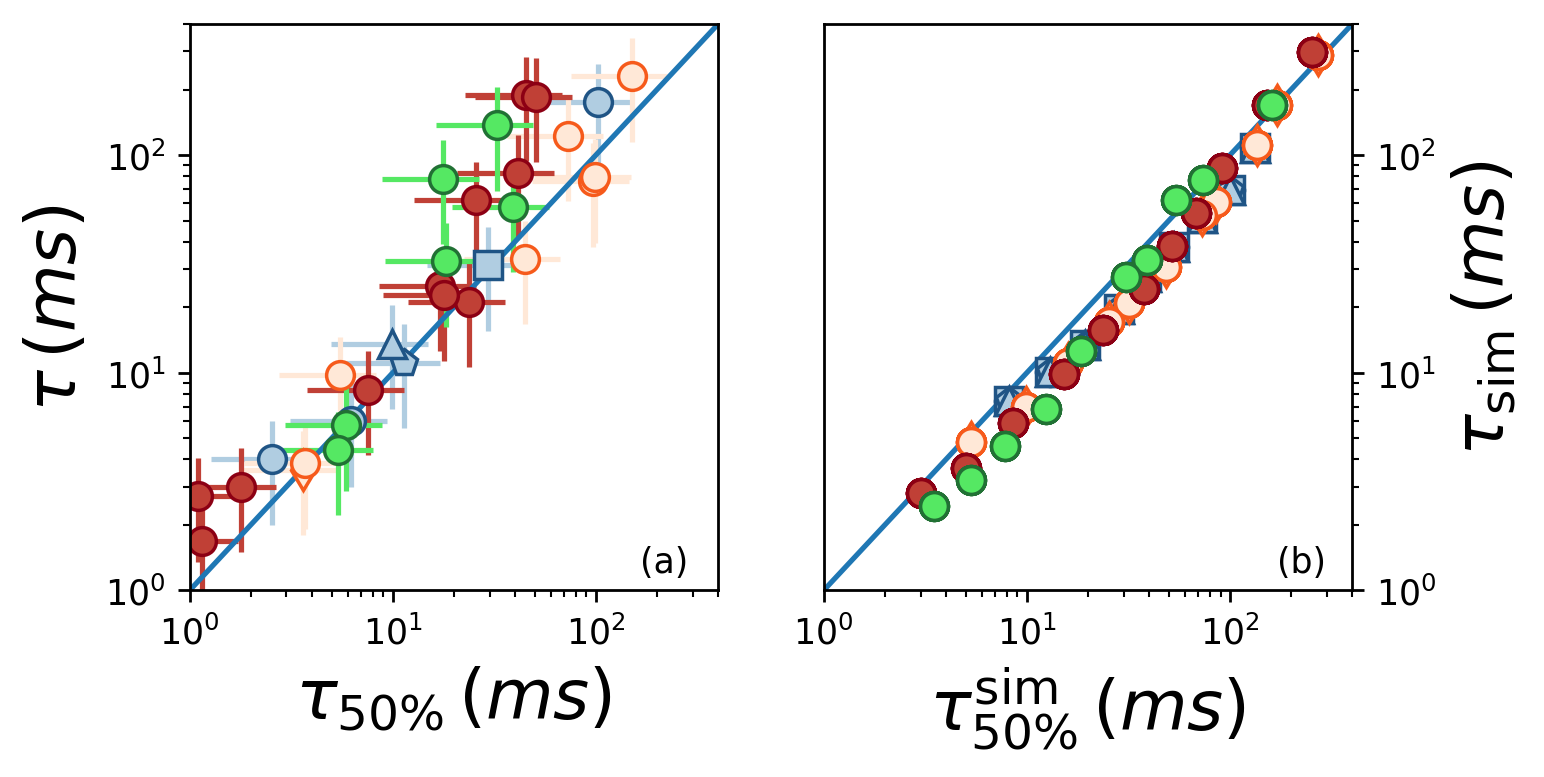}
    \caption{$\tau$ vs \taub for experimental (a) and numerical (b) results. The two quantity are compatible to each other within error bars in experimental results and show only small deviation in simulation data. Color code for the points is the same used in the main text and the blue line marks $\tau=$\taub. }
    \label{S3}
\end{figure}

To verify the solidity of our approach we also considered a different characteristic time \taub defined as the time at which the probability $P$ to observe a bubble is $50\%$. This approach is a valid alternative for measurements featuring a limited statistics.
To determine \taub we fit $P$ with the following function:
\begin{equation}
    P(t)=\mathrm{Min}[a_1*(e^{t/a_2}-1),1]
\end{equation}
with $a_1$ and $a_2$ as free parameters. These two are then used to determine \taub from
\begin{equation}
    \frac{1}{2}=a_1*(e^{t_{\mathrm{50\%}}/a_2}-1)
\end{equation}
We check, within the statistical uncertainties, that the value of \taub does not change by using different fitting functions (linear, exponential with offsets in time and $P$). Figure \,\ref{S3} shows that $\tau$ and \taub are compatible both for the experimental measurements and numerical simulations. In particular, simulation results allow us to conclude that, while \taub is expected to be influenced by the delay time before the bubble decays, \taub is still a good approximation of $\tau$. This suggests that the delay time and $\tau$ are related and further investigations are necessary to understand how.

In general, we conclude that the determination of $\tau$ used in the main text is solid. In particular, one notes that the two methods rely on two very different observables, the mean magnetization in the center, averaged over all experimental shots ($\tau$), and the probabilistic presence of a bubble (\taub).

\section{Numerical simulations}
\label{simulation}

The numerical results presented in the main text are based on one-dimensional Gross-Pitaevskii simulations. The parameters are chosen to faithfully reproduce the experimental conditions: in particular, the system trapped by a harmonic potential with frequency $\omega_0 \simeq 2\pi\times$16~Hz, so that the Thomas-Fermi radius is $L \simeq 200\,\mu $m; moreover, interactions are chosen to obtain $|\kappa| n_0 = |\Delta| n_0 \simeq 2\pi \times$1.1~kHz, $n_0$ being the total density in the center of the cloud. 
The system is first prepared, through imaginary-time evolution, in the ground state corresponding to $\delta_{f} = 2\pi \times$1~kHz, thus, regardless of the value of $\Omega_R$, it is almost fully polarized in the $|\uparrow\rangle$ state. 

A white noise of amplitude equal to $3\%$ of the central density is added on top of the ground state: this corresponds to an injected energy of roughly $\varepsilon/k_B = 215~\text{nK}$. We then let the system evolve in real time, without changing any parameter and we observe that, after a transient, the noise distribution becomes stationary; we interpret this result as thermalization of the mixture to a temperature $T\propto \varepsilon$. Under an ergodicity assumption, we can determine the dynamics of the system by averaging over many repetitions of the same time-evolution, each one obtained starting from a different noisy sample. To summarize, we perform mean-field simulations in which noise plays the role of an effective temperature. Of course, these do not allow to investigate the role of quantum fluctuations: however, since the estimated experimental temperature is much higher than $|\kappa|n_0/k_B \sim 50 \text{nK}$, the dynamics is likely to be dominated by thermal noise and a comparison with classical field simulations is justified. 

The real-time dynamics after thermalization reproduces, once again, the experimental protocol: a detuning ramp with speed $\sim 50~ \text{Hz/ms}$ is applied in order to reach the false vacuum state corresponding to some final $\delta_f < 0$; the magnetization of the system is then monitored for a waiting time in the range $[10, 300]\, \text{ms}
$, depending on the simulation parameters. 

In order to extract the characteristic decay time $\tau$ and $\tau_{50}$, we compute:
\begin{equation}
    F(t) = \frac{\langle Z(x\sim 0, t)\rangle -Z_{TV}}{Z_{FV}-Z_{TV}} 
\end{equation}
where $\langle Z(x\sim 0, t)\rangle$ is the statistical average of magnetization over the central $10~\mu$m of the cloud. If the number of samples is sufficiently high (we use 1000), this function represents the probability of not observing a bubble at time $t$. Therefore, $\tau_{50}$ is computed, by definition, by solving $F(\tau_{50}) = 0.5$. 

The FVD rates are obtained instead via a linear fit of $\log F(t)$: in most cases the predicted exponential behaviour is found within a time interval corresponding to $F(t)\in [0.3, 0.7]$; small adjustments of this window are necessary for the simulations associated to the smallest and longest tunnelling times. 

\section{Istantons}

The theoretical description of vacuum decay is non-perturbative and
based on instanton solutions to the equations of motion using an imaginary time coordinate. 
The classical field theory for this system reduces down to a field theory for the magnetisation $Z$. For thermal instantons, bubbles nucleate at a rate (see e.g.\cite{Hindmarsh2021})
\begin{equation}
\Gamma=1/\tau=A\left(\beta E_c\right)^{j/2}e^{-\beta E_c}.
\label{ThermalRate}
\end{equation}
where $\beta=1/(k_BT)$ and $E_c$ is the energy of the instanton. The factor $A$ depends on fluctuations about the instanton and $j$ is the number of translational symmetries.
There should be one zero mode $j=1$ if there is translational invariance
in the system. (The bubbles in the experiment always nucleate near the centre, so translational invariance is suspect. Fortunately, the power law dependence has only a small effect on the results). There are a very limited number of  models for which the pre-factor $A$ is calculable at present, and we will therefore regard $A$ as a fitting parameter in the subsequent analysis. Note that the non-perturbative approach is valid when the exponent is larger than one, i.e. for temperatures $k_BT<E_c$. At even lower temperatures, vacuum fluctuations become the dominant seeding mechanism. In our
system this happens for $k_BT<\hbar |\kappa|n\sim 50~{\rm nK}$, and the resulting vacuum decay rate would be far less than the rate seen in the experiment.

The energy for a thermal instanton includes a gradient contribution
\begin{equation}
E_c=\frac{\hbar n}{4}\int\left\{
\frac{\hbar}{2m}\frac{(\nabla Z)^2}{1-Z^2}
+V\right\}dx,\label{Action}
\end{equation}
where the potential
\begin{equation}
V=\kappa n Z^2-2\Omega_R(1-Z^2)^{1/2}-2\delta_{\rm f} Z.
\end{equation}
We can scale out the dependence on the
density so that $\hat E_c=E_c/(\hbar n^2\xi|\kappa|)$ for the length scale
$\xi=\hbar/(m|\kappa|n)^{1/2}$. For thermal bubbles in one dimension, the 
instanton calculation is equivalent to a WKB 
approximation to the action, with the familiar WKB form
\begin{equation}
\hat E_c=\frac12\int_{Z_{TP}}^{Z_{FV}}\left(\frac{2(V-V_{FV})}{|\kappa|n}\right)^{1/2}\frac{dZ}{\sqrt{1-Z^2}},
\label{ThermalAction}
\end{equation}
The integral extends from the turning point $Z_{TP}$ to the false vacuum $Z_{FV}$.
The extra factor $(1-Z^2)^{-1/2}$ is due to the form of the derivative terms in the energy (\ref{Action}).

The experimental data has been used to determine the best parameters in a fit for 
$\ln\tau=\ln A+b\hat E_c-\ln(b\hat E_c)/2$. The results are given in \aref{tab:fits}. 
The condensate number density is given by $n=(k_BT/\hbar|\kappa|n)b/\xi$. For the temperature $T=1~\mu$K, the values of $n$ at lower $\Omega$ are around half of the value expected for the system, but not unreasonable given the limitations of the one dimensional treatment. If the bubble only fills a fraction of the cross-section, it effectively feels only part of the integrated density.

\begin{table}[]
\caption{\label{tab:fits}Fitting coefficients for the thermal instanton model of vacuum decay with $j=1$. The fit is limited to $(\delta_f-\delta_{\rm c})/\Omega_R>0.05$ to ensure that $b\hat E_c>1$.}
\begin{ruledtabular}
\begin{tabular}{llllll}
$\Omega_R/2\pi$ & $a_\text{exp}(\sigma_a)$& $b_\text{exp}(\sigma_b)$& $a_\text{sim}(\sigma_a)$& $b_\text{sim}(\sigma_b)$\\
\hline  \\ [-1.5ex]
300&0.54(0.09)&56.5(1.9)&0.93(0.06)&55.0(1.9)\\
400&0.83(0.42)&44.4(6.1)&0.70(0.07)&41.3(0.87)\\
600&0.02(0.43)&30.3(3.7)&0.01(0.14)&29.8(1.3)\\
800&0.30(0.75)&25.8(5.7)&-0.44(0.11)&25.3(0.9)\\
\end{tabular}
\end{ruledtabular}
\end{table}

\begin{figure}[t!]
    \centering
    \includegraphics[width = .55\columnwidth]{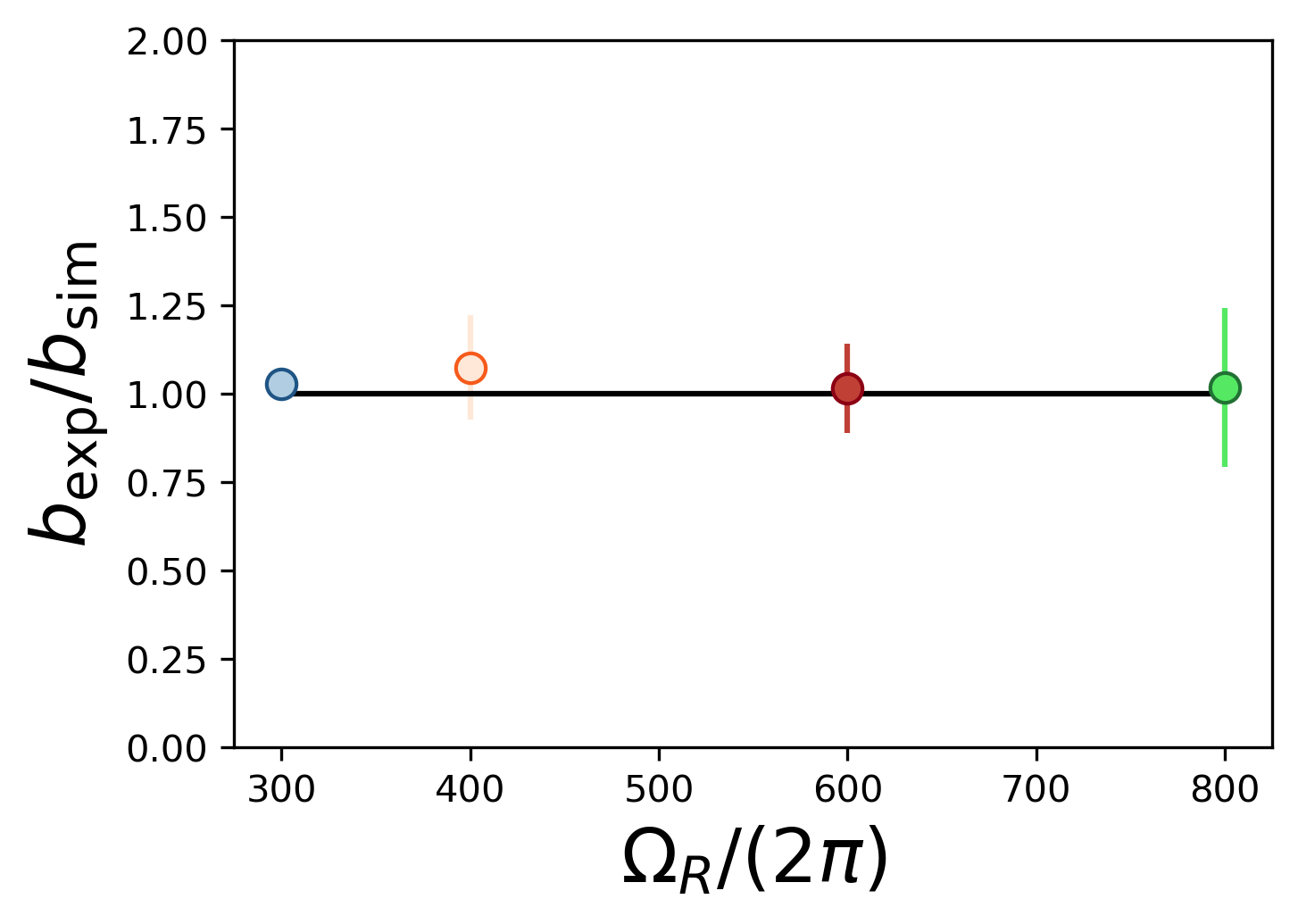}
    \caption{Ratio between of $b$ parameters from instanton fit on the experiments and simulations. Error bars results from the statistical uncertainties on the fits.}
    \label{S4}
\end{figure}

In the case of small potential barriers, the potential can be expanded to cubic order about an inflection
point at $Z_c$ and $\delta=\delta_c$, where
\begin{equation}
\delta_c=\kappa n(1-Z_c^3),\qquad 
Z_c=\left(1-\left(\frac{\Omega_R}{|\kappa|n}\right)^{\textstyle\frac23}
\right)^{\textstyle\frac12}.
\end{equation}
The integral in this case can be performed exactly,
\begin{equation}
\hat E_c\approx1.77\left(\frac{\delta_f-\delta_c}{|\kappa|n}\right)^{\textstyle\frac54}
\left(\frac{\Omega_R}{|\kappa|n}\right)^{\textstyle\frac{1}{6}}
\left(\frac{|\delta_c|}{|\kappa|n}\right)^{\textstyle-\frac14}
\end{equation}

To verify that the instanton prediction and simulation are consistent, we repeat numerical simulations at fixed $\delta_f$ and variable $\varepsilon$. We observe that the extracted $\tau$ results proportional to $e^{(1/\varepsilon)}$ and this well justifies the association between the injected noise parameter $\varepsilon$ and the temperature $T$.

\section{Pseudo-relativistic behaviour}

In the present work we did not need to consider the dynamical properties of the field, but the thermally driven tunneling rate. It is however worth mentioning that coherently coupled Bose-Einstein condensates can show pseudo-relativistic behaviour at small momenta. The dispersion relation of the spin channel as long as the coupling with the density channel can be neglected reads \begin{equation}
   \omega^2 = \left(\frac{\hbar k^2}{2m}+ {\Omega_R\over\sqrt{1-Z^2}} \right)
\left(\frac{\hbar k^2}{2m}+ {\Omega_R\over\sqrt{1-Z^2}} +(1-Z^2) \kappa n\right)
\end{equation}
Therefore in the small momenta regime one has pseudo-relativistic regions, where $\omega^2\propto k^2 + M^2$, reinforcing the analogy with early universe phase transitions, which occur in a relativistic plasma. 
Moreover the dispersion relation of the system's excitation and more generally the time and space derivatives entering in the effective field theory play an important role in the bubble dynamics after its formation \cite{Coleman1977a}. 

\section{Determination of $\sigma_\mathrm{f}$}

\begin{figure}[b]
    \centering
    \includegraphics[width = .50\columnwidth]{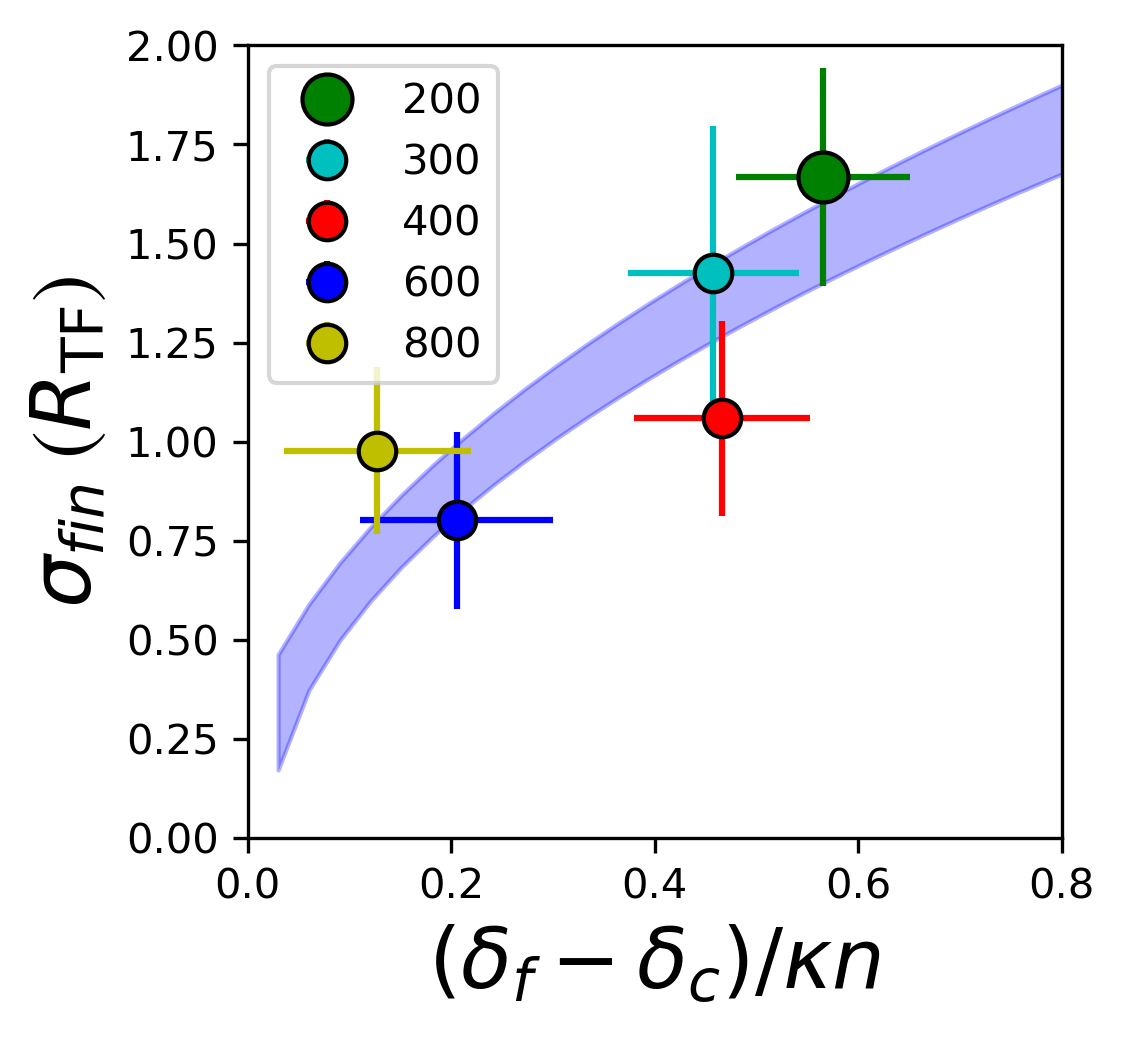}
    \caption{Size of the bubble once the stationary state is reached after a long waiting time. The points of different colour  indicate the experimental measurements at various $\Omega_R/2\pi$, and they are in good agreement with the size expected from the numerical model (shaded area). }
    \label{S5}
\end{figure}

The measured size of the bubble increases in time and saturates to a final value. This suggests that the highly energetic resonant state populated after the tunneling process dissipates energy and the size of bubble reaches a steady value. Despite we are not able to follow the time evolution of the single bubble but only the averaged size after a variable time from its probabilistic formation, we verify that the final size of the bubble corresponds to the size of the magnetic ground state; see Fig.\,\ref{S5}.
The theoretical final size of the bubble is determined by the size of the ferromagnetic region of the cloud. Due to atom losses and finite coherence time of the coupled mixture, we can not wait an indefinitely long time. To overcome this limitation and extract $\sigma_\mathrm{f}$ as the asymptotic value of the experimental sizes. We evaluate the averaged bubble size $\langle \sigma \rangle$ at time $t$, by fixing $\sigma =0$ in absence of bubbles. We then fit it with the sigmoidal function 
\begin{equation}
f    \langle \sigma (t) \rangle = \sigma_\mathrm{f} \mathrm{Max}\left[ 0,\arctan \left (\frac{(t-t_0)}{\tau_{\sigma}} \right ) \right ]
\end{equation}
with $\sigma_\mathrm{f}$, $t_0$ and $\tau_{\sigma}$ as free parameters.

\end{document}